\documentclass[aps,prl,twocolumn,superscriptaddress]{revtex4-1}

\usepackage{graphicx}
\usepackage{dcolumn}
\usepackage{bm}
\usepackage{hyperref}
\usepackage{todonotes}
\usepackage{color}
\usepackage{glossaries}

\newacronym{RIXS}{RIXS}{resonant inelastic x-ray scattering}
\newacronym{FWHM}{FWHM}{full-width at half-maximum}
\newacronym{QCP}{QCP}{quantum critical point}
\newacronym{RPA}{RPA}{random phase approximation}
\newacronym{SOC}{SOC}{spin-orbit coupling}
\newacronym{SI}{SI}{Supplementary Information}
\begin{document}

\title{Antiferromagnetic Excitonic Insulator State in Sr$_3$Ir$_2$O$_7$}

\author{D. G. Mazzone}
\email{daniel.mazzone@psi.ch}
\affiliation{Department of Condensed Matter Physics and Materials Science, Brookhaven National Laboratory, Upton, New York 11973, USA}
\affiliation{Laboratory for Neutron Scattering and Imaging, Paul Scherrer Institut, CH-5232 Villigen, Switzerland}

\author{Y. Shen}
\affiliation{Department of Condensed Matter Physics and Materials Science, Brookhaven National Laboratory, Upton, New York 11973, USA}

\author{H. Suwa}
\affiliation{Department of Physics, The University of Tokyo, Tokyo 113-0033, Japan}
\affiliation{Department of Physics and Astronomy, University of Tennessee, Knoxville, Tennessee 37996, USA}

\author{G. Fabbris}
\affiliation{Advanced Photon Source, Argonne National Laboratory, Argonne, Illinois 60439, USA}

\author{J. Yang}
\affiliation{Department of Physics and Astronomy, University of Tennessee, Knoxville, Tennessee 37996, USA}

\author{S-S. Zhang}
\affiliation{Department of Physics and Astronomy, University of Tennessee, Knoxville, Tennessee 37996, USA}

\author{H. Miao}
\affiliation{Material Science and Technology Division, Oak Ridge National Laboratory, Oak Ridge, Tennessee 37831, USA}

\author{J. Sears}
\affiliation{Department of Condensed Matter Physics and Materials Science, Brookhaven National Laboratory, Upton, New York 11973, USA}

\author{Ke Jia}
\author{Y. G. Shi}
\affiliation{Beijing National Laboratory for Condensed Matter Physics, Institute of Physics, Chinese Academy of Sciences, Beijing 100190, China}

\author{M. H. Upton}
\author{D. M. Casa}
\affiliation{Advanced Photon Source, Argonne National Laboratory,
Argonne, Illinois 60439, USA}

\author{X. Liu}
\email{liuxr@shanghaitec.edu.cn}
\affiliation{School of Physical Science and Technology, ShanghaiTech University, Shanghai 201210, China}

\author{J. Liu}
\affiliation{Department of Physics and Astronomy, University of Tennessee, Knoxville, Tennessee 37996, USA}

\author{C. D. Batista}
\affiliation{Department of Physics and Astronomy, University of Tennessee, Knoxville, Tennessee 37996, USA}
\affiliation{Quantum Condensed Matter Division and Shull-Wollan Center, Oak Ridge National Laboratory, Oak Ridge, TN 37831, USA}

\author{M. P. M. Dean}
\email{mdean@bnl.gov}
\affiliation{Department of Condensed Matter Physics and Materials Science, Brookhaven National Laboratory, Upton, New York 11973, USA}

\def\mathbi#1{\ensuremath{\textbf{\em #1}}}
\def\Q{\ensuremath{\mathbi{Q}}}
\def\SIO{Sr$_3$Ir$_2$O$_7$}
\newcommand{\angstrom}{\mbox{\normalfont\AA}}
\date{\today}

\begin{abstract}
Excitonic insulators are usually considered to form via the condensation of a soft charge mode of bound electron-hole pairs. This, however, presumes that the soft exciton is of spin-singlet character. Early theoretical considerations have also predicted a very distinct scenario, in which the condensation of magnetic excitons results in an antiferromagnetic excitonic insulator state. Here we report \gls*{RIXS} measurements of Sr$_3$Ir$_2$O$_7$. By isolating the longitudinal component of the spectra, we identify a magnetic mode that is well-defined at the magnetic and structural Brillouin zone centers, but which merges with the electronic continuum in between these high-symmetry points and which decays upon heating concurrent with a decrease in the material’s resistivity. We show that a bilayer Hubbard model, in which electron-hole pairs are bound by exchange interactions, consistently explains all the electronic and magnetic properties of Sr$_3$Ir$_2$O$_7$ indicating that this material is a realization of the long-predicted antiferromagnetic excitonic insulators phase.
\end{abstract}

\maketitle

\section*{Introduction}
Detailed theoretical considerations of narrow-gap insulators date back to the 1960s, when it was realized that if the energy required to form an electron-hole pair becomes negative, a phase transition into an excitonic insulator state can occur \cite{Mott1961transition,Keldysh1965possible, Jerome1967excitonic, Halperin1968possible}. Unscreened electron-hole Coulomb attraction is perhaps the most obvious driving force behind this phase transition, and excitonic charge insulator states are indeed thought to occur in materials such as TmSe$_{0.45}$Te$_{0.55}$, 1T‐TiSe$_2$, and Ta$_2$NiSe$_5$ \cite{Bucher1991excitonic, Wakisaka2009excitonic, Eisenstein2014exciton, Kogar2017signatures, Lu2017zero}. Although less intuitive, effective electron-hole attraction can also arise from on-site electron-electron Coulomb repulsion $U$ via magnetic exchange interactions between the electron and hole \cite{Slater1951magnetic}. In this case, the soft exciton is expected to be a spin-triplet, which passes through a \acrfull*{QCP} with increasing effective $U$. The condensation of the relevant triplet exciton at the \gls*{QCP} gives rise to an antiferromagnetic ground state hosting a well-defined excitonic longitudinal mode \cite{Halperin1968possible}, which coexists with transverse modes that are a generic feature of ordered antiferromagnets. This longitudinal mode features excitonic character, in the sense that it modifies the local spin amplitude by creating electron-hole pairs \cite{Halperin1968possible}. In this work we identify and study a longitudinal mode in Sr$_3$Ir$_2$O$_7$, the presence of which is the key experimental signature of an antiferromagnetic excitonic insulator.

\section*{Results}

\begin{figure*}
\includegraphics[width=0.72\linewidth]{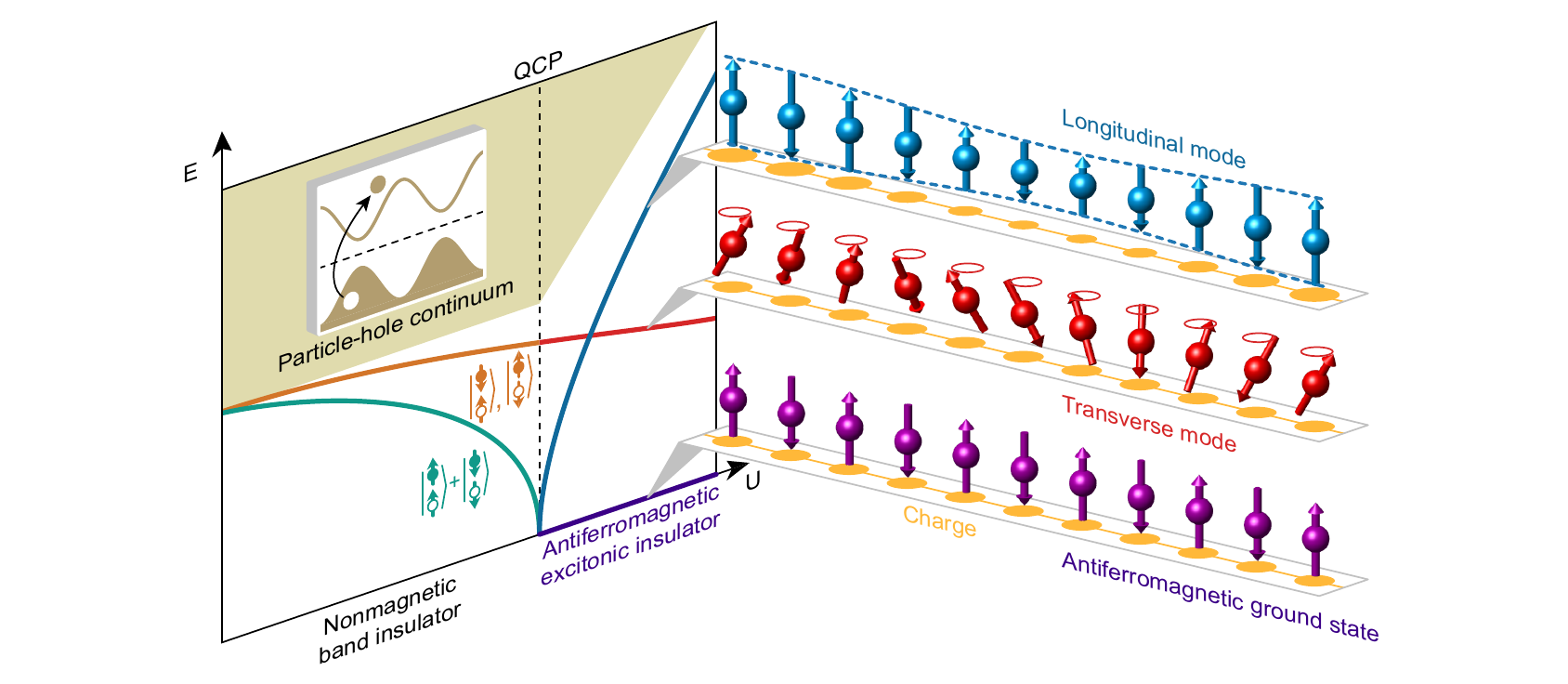} %
\caption{\textbf{Antiferromagnetic excitonic insulator phase diagram.}
Charge excitations in paramagnetic band insulators consist of either electron-hole excitations across the insulating band gap (brown shaded area) or of bound electron-hole excitons below the particle-hole continuum [electrons (holes) are indicated with filled (empty) circles]. An antiferromagnetic excitonic insulator is established through the condensation of the predominately spin-triplet character exciton mode with spin quantum number $S^z=0$. The excition is a superposition of an up-spin electron in the conduction band paired with an up-spin hole (equivalent to a down-spin electron) and a down-spin electron paired with an down-spin hole \cite{Halperin1968possible}. The other spin-triplet excitions $S^z=\pm1$ feature an up-spin electron and a down-spin hole or a down-spin electron and an up-spin hole. Upon increasing Coulomb interaction $U$ the $S^z=0$ exciton condenses into the ground state at a \gls*{QCP} \cite{Halperin1968possible}, establishing magnetic order and leaving an excitonic longitudinal mode as the key signature of this state.}
\label{fig:schematic}
\end{figure*}

\begin{figure}
\includegraphics[width=0.92\linewidth]{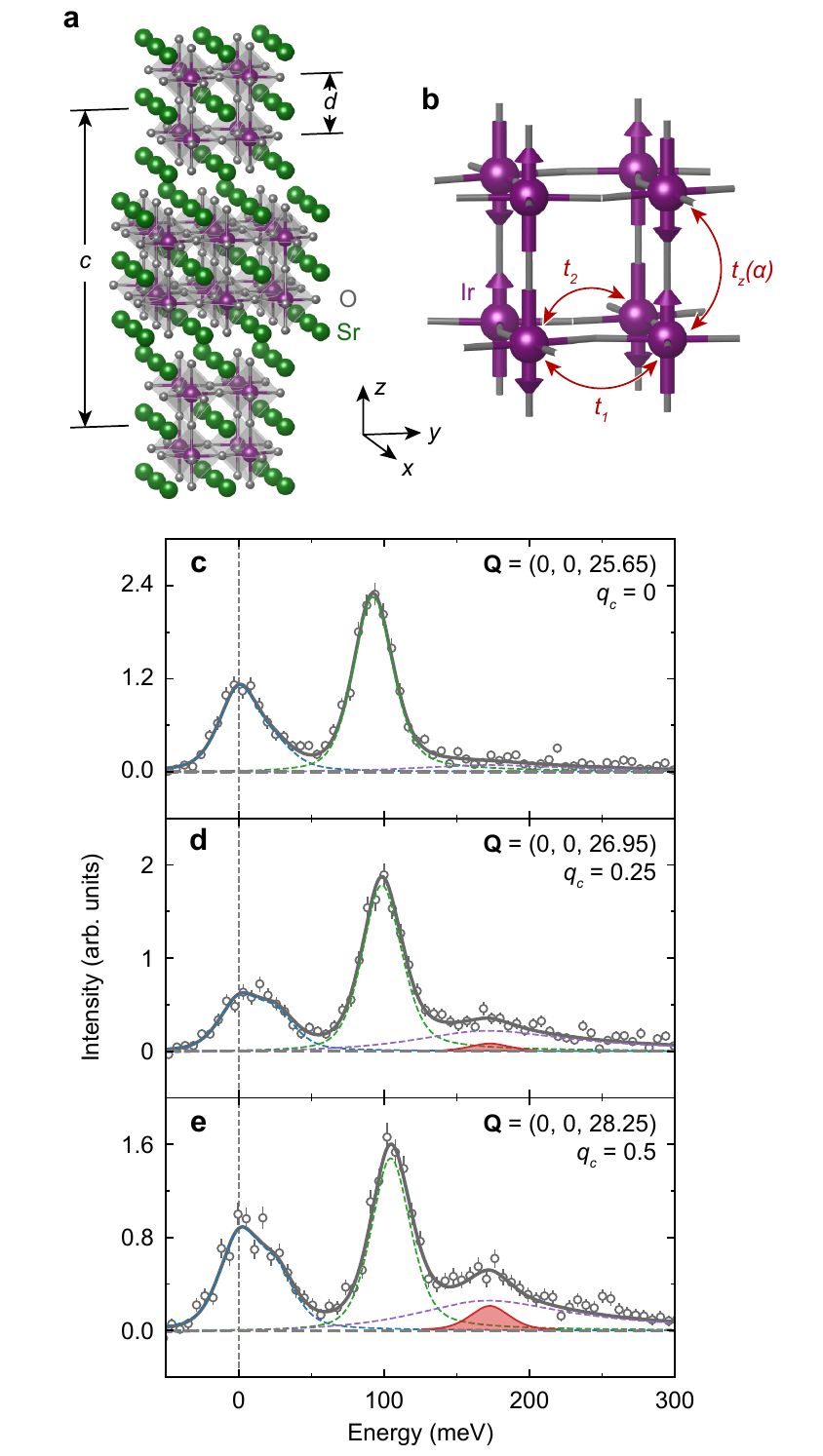} %
\caption{\textbf{Isolating the excitonic longitudinal mode in Sr$_3$Ir$_2$O$_7$.}
\textbf{a}  Crystal structure of the bilayer material Sr$_3$Ir$_2$O$_7$. \textbf{b} Ir-Ir bilayer with $t_1$ the nearest-neighbour, $t_2$ the next-nearest-neighbour and  $t_z(\alpha)$ the interlayer hopping terms. \textbf{c-e} \Gls*{RIXS} spectra measured at $T = 20$~K and $\mathbf{Q} = (0, 0, L)$ with $L = 25.65, 26.95$ and 28.25 in reciprocal lattice units. The $c$-axis positions are also labelled in terms of the Ir-Ir interlayer reciprocal-lattice spacing $q_c = 0, 0.25$ and 0.5.  An additional mode appears around 170~meV with maximal intensity at $q_c = 0.5$ (see shaded red area). The black circles represent the data and dotted lines outline the different components of the spectrum, which are summed to produce the grey line representing the total spectrum.  Error bars are determined via Poissonian statistics.}
\label{fig:ldep}
\end{figure}

The formation of an antiferromagnetic excitonic  insulator requires a very specific set of conditions. We need (i) a charge gap of similar magnitude to its magnetic energy-scale and (ii) strong easy-axis anisotropy. Property (i) is a sign that the material is close to the excitonic \gls*{QCP} (see Fig.~\ref{fig:schematic}). Property (ii) is not a strict condition, but it facilitates the identification of an antiferromagnetic excitonic insulator, because the opening of a spin gap $\Delta_s$ protects the longitudinal mode from decay. This is because longitudinal fluctuations are often kinetically predisposed to decay into transverse modes generating a longitudinal continuum with no well-defined modes. This decay can be avoided when the energy of the longitudinal mode is lower than twice the spin gap. Iridates host strong \gls*{SOC}, which can help realize a large spin gap and bilayer Sr$_3$Ir$_2$O$_7$ shown in Fig.~\ref{fig:ldep}a is known to have a narrow charge gap of order $\Delta_c\sim 150$~meV \cite{Moon2008}. The essential magnetic unit, with $c$-axis ordered moments, is shown in Fig.~\ref{fig:ldep}b \cite{Kim2012_2}. In view of the antiferromagnetic order in Sr$_3$Ir$_2$O$_7$, the material would be predicted to lie in the magnetically ordered region  to the right of the \gls*{QCP}  where the excitonic longitudinal mode is expected to appear. Because the exciton is predicted to have odd parity under exchange of the two Ir layers, we expect the excitonic longitudinal mode to be present at $c$-axis wavevectors corresponding to antisymmetric bilayer contributions and absent at the symmetric condition. We label these wavevectors $q_\text{c}=0.5$ and $q_\text{c}=0$, respectively. In contrast, transverse magnetic modes are expected to be present at all $c$-axis wavevectors, allowing the transverse and longitudinal modes to be readily distinguished.

The excitation spectrum of Sr$_3$Ir$_2$O$_7$ was studied with \gls*{RIXS}. Figures \ref{fig:ldep}c-e display energy-loss spectra at $T = 20$~K, well below the N\'{e}el temperature $T_\text{N}= 285$~K and $q_\text{c} = 0, 0.25$ and 0.5, corresponding to $L = 25.65, 26.95$ and 28.25 in reciprocal lattice units (r.l.u.). These irrational $L$ values arise because the bilayer separation $d$ is not a rational fraction of the unit cell height $c$ (see Methods section for details). The spectrum at $q_c = 0$ is composed of a phonon-decorated quasi-elastic feature, a pronounced magnetic excitation at $\sim100$~meV which we later identify as the transverse mode, and a high-energy continuum. As explained above, changing $q_c$ is expected to isolate the anticipated excitonic mode. A longitudinal mode is indeed observed, reaching maximum intensity at $q_c = 0.5$, and is highlighted by red shading in Figs.~\ref{fig:ldep}d and e.

\begin{figure*}
\includegraphics[width=0.92\linewidth]{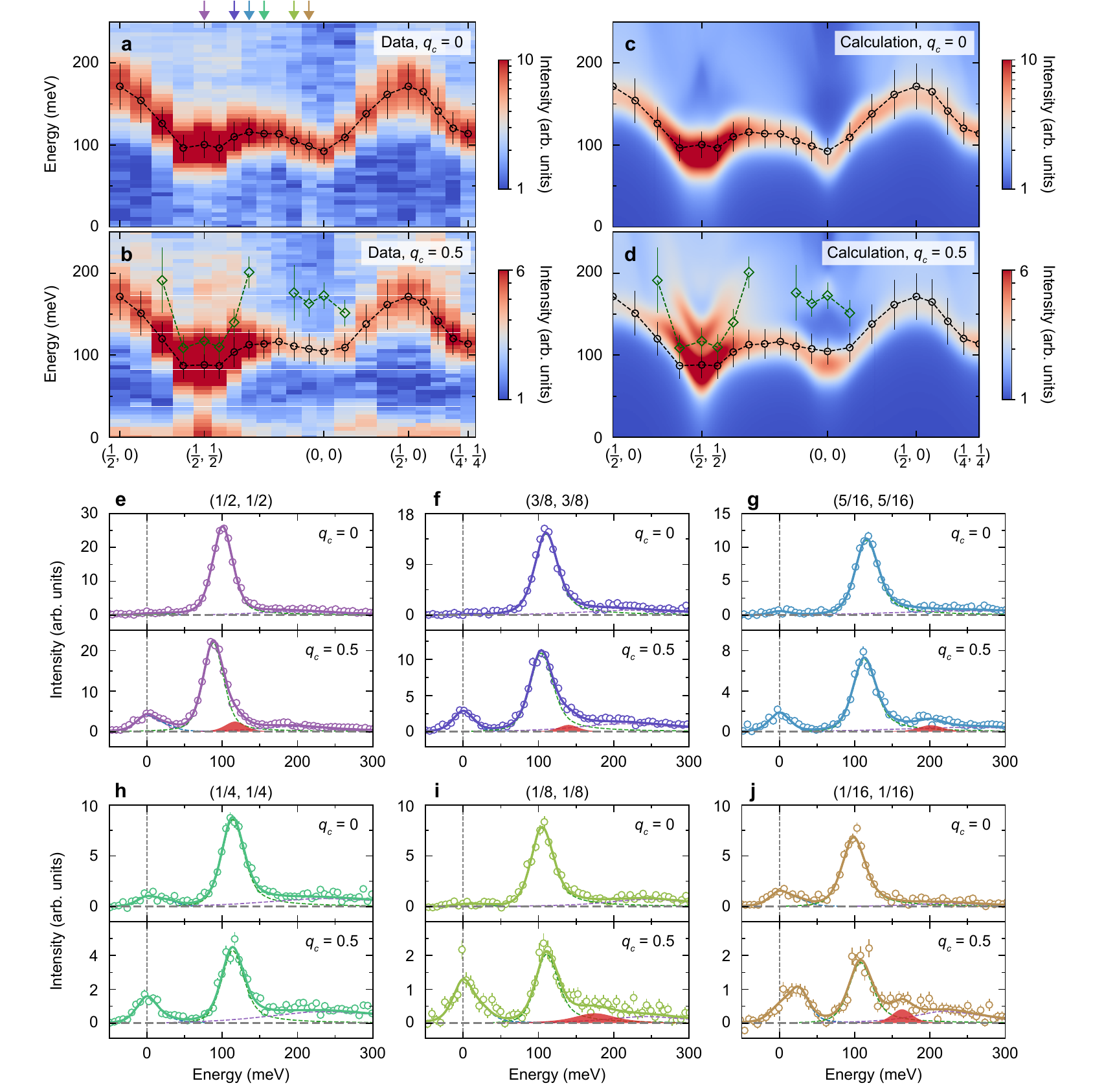} %
\caption{\textbf{Magnetic dispersion and excitonic longitudinal mode decay.} \textbf{a} and \textbf{b} In-plane momentum dependence of the magnetic excitations measured at $q_c = 0$ and 0.5. The black and green symbols correspond to the energy of the magnetic modes and the vertical bars to their peak widths. Both quantities were extracted from the energy spectra at different points in reciprocal space (such as shown in panels \textbf{e}-\textbf{j} and Fig.~\ref{fig:ldep}c-e). \textbf{c} and \textbf{d} Theoretical calculations of the magnetic dispersion relation, overplotted with the experimentally determined excitation energies and line widths. The presence of the mode at $q_c = 0.5$ that is absent at $q_c = 0$ evinces that this is an excitonic longitudinal mode.  \textbf{e-j} \Gls*{RIXS} spectra at reciprocal space as highlighted by color-matching arrows in panel \textbf{a}. Circles represent the data and dotted lines outline the different components of the spectrum, which are summed to produce the solid line representing the total spectrum. Error bars are determined via Poissonian statistics. The isolation of the longitudinal mode (highlighted with red shading) from other contributions was possible by simultaneously analyzing $q_c = 0.5$ and $q_c = 0$ for each in-plane reciprocal-lattice wavevector (see Methods section for details).}
\label{fig:dispersion}
\end{figure*}

In isolation, the presence of a longitudinal magnetic mode in this symmetry channel is a necessary but insufficient condition to establish an antiferromagnetic excitonic insulator, so we leverage the specific symmetry, decay, and temperature dependence of the longitudinal and transverse magnetic modes to establish the presence of the novel state. The only other candidate magnetic model that hosts a longitudinal mode of this type is a specific configuration of the bilayer Heisenberg Hamiltonian, in which the charge degrees of freedom are projected out. In particular, a model with a $c$-axis magnetic exchange $J_c$ that is larger than, but not dramatically larger than, the in-plane exchange $J_{ab}$ is needed to produce a longitudinal mode and large easy-axis magnetic anisotropy is required to reproduce the spin gap. If $J_c \ll J_{ab}$, the spectrum would show only a spin-wave-like in-plane dispersion  contrary to the observed $q_c$ dependence in Fig.~\ref{fig:ldep}c-e, and in the $J_c \gg J_{ab}$ limit the system would become a quantum paramagnet. For $J_c / J_{ab}$ of order two, the bilayer Heisenberg Hamiltonian supports a longitudinal mode and for the current case of large easy-axis anisotropy, the transverse and longitudinal modes  appear as well-defined modes throughout the Brillouin zone \cite{Lohofer2015dynamical, Moretti2015evidence, Zhou2020amplitude, Su2020stable}. In fact, earlier reports have proposed this spin dimer model to explain \gls*{RIXS} measurements of the longitudinal $\sim170$~meV feature in Sr$_3$Ir$_2$O$_7$ \cite{Moretti2015evidence, Hogan2016disordered}. Although prior and subsequent non-dimerized models have also been proposed to describe Sr$_3$Ir$_2$O$_7$ as rival candidates \cite{Kim2012_2, Gretarsson2016two, Lu2017doping, Li2020symmetry, Mohapatra2017}. These models, however, do not support a longitudinal mode (a detailed comparison between the different models is given in \gls*{SI} Section~1). 
We therefore map the in-plane dispersion relations at $q_c = 0$ and 0.5 and show them in Figs.~\ref{fig:dispersion}a,b. At $q_c = 0$, where the longitudinal mode is suppressed by symmetry, we observe an excitation dispersing from $\sim90$ to 170~meV and a continuum at higher energies. Simultaneously analyzing $q_c = 0.5$ and $q_c = 0$ for each in-plane reciprocal-lattice wavevector, while leveraging the distinct symmetry properties of the longitudinal and transverse modes, allows us to isolate the longitudinal mode (see Methods section). We plot the position and peak width of the longitudinal mode in green on Fig.~\ref{fig:dispersion}b. The transverse mode, on the other hand, is symmetry-allowed at $q_c = 0.5$ and $q_c = 0$ and is shown in black on Fig.~\ref{fig:dispersion}a,b. We find that the longitudinal mode is well-defined around $(0, 0)$ (Fig.~\ref{fig:ldep}c-d and Fig.~\ref{fig:dispersion}i), but decays into the high-energy continuum as it disperses away, becoming undetectable at $(1/4, 1/4)$ (Fig.~\ref{fig:dispersion},h). The longitudinal mode is also detectable as a shoulder feature on the transverse mode at $(1/2, 1/2)$ before dispersing upwards and broadening at neighboring momenta (Fig.~\ref{fig:dispersion}e, g). The decay and merging of the longitudinal mode into the electron-hole continuum was not detected previously and suggests the realization of an antiferromagnetic excitonic insulator state because the longitudinal mode in this model has a bound electron-hole pair character and therefore will necessarily decay when it overlaps with the electron-hole continuum. This longitudinal mode decay is incompatible with a longitudinal mode arising from spin-dimer excitations in a strongly isotropic bilayer Heisenberg model, which predicts well-defined modes throughout the Brillouin zone and projects out the high-energy particle-hole continuum~\cite{Lohofer2015dynamical, Moretti2015evidence, Zhou2020amplitude, Su2020stable}.

\begin{figure}
\includegraphics[width=0.92\linewidth]{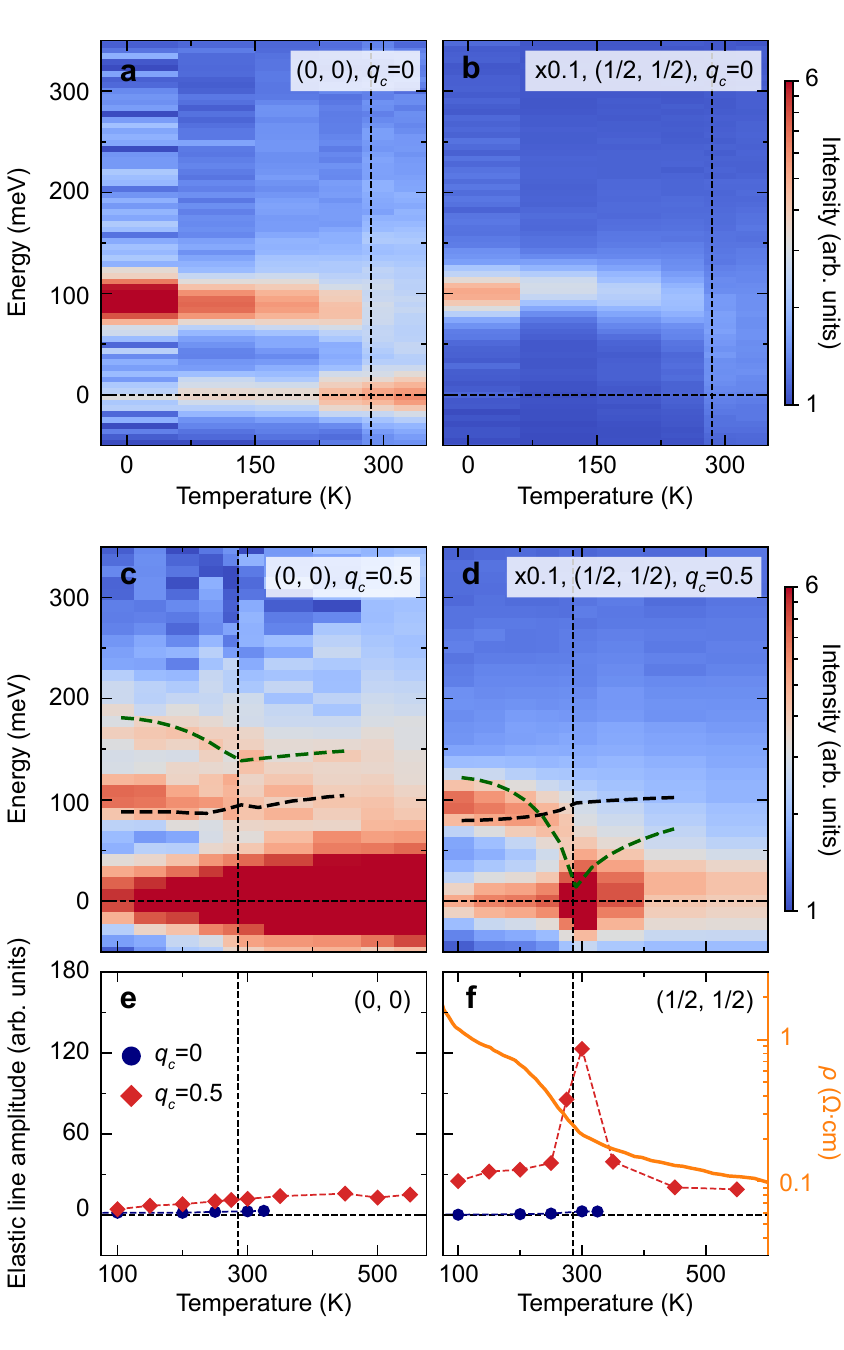} %
\caption{\textbf{Excitonic mode condensation at the N\'{e}el temperature.} \textbf{a-d} Temperature dependence of the Sr$_3$Ir$_2$O$_7$ excitation spectrum at (0, 0) and $(0.5, 0.5)$  for $q_c = 0$ and 0.5 \Gls*{RIXS} spectra at selected temperatures are shown in Fig.~S2). The intensity at $(0.5, 0.5)$ has been scaled for comparison reasons. The dashed lines show temperature-dependent calculations of our model (the full theoretical predictions are plotted in Fig.~S5). Based on the $q_c$ behavior of the modes, we know that panels \textbf{a, b} show only the transverse mode, while \textbf{c, d} show both the transverse and longitudinal mode.  \textbf{e, f} Quasi-elastic intensity as function of temperature for $q_c = 0$ and 0.5 in blue and red, respectively. The non-monotonic enhancement at $q_c = 0.5$ in \textbf{e} provides additional support that the condensation of the excitonic longitudinal mode establishes the magnetic long-range order in Sr$_3$Ir$_2$O$_7.$ Panel \textbf{f} also shows the  anomalous temperature dependence of the electric resisitivty $\rho$ (taken from \cite{Cao2002anomalous}), which shows a change in gradient at $T_\text{N}$ further indicating that charge fluctuations are involved in the transition.}
\label{fig:tdep}
\end{figure}

Since optical conductivity, tunneling spectroscopy, and photo-emission studies all report charge gaps $\Delta_c$ on the same energy-scale of magnetic excitations (100-200~meV) \cite{Moon2008, Okada2013imaging, Wang2013, King2013spectroscopic, Torre2014coherent}, we model the microscopic interactions within a Hubbard Hamiltonian that retains the charge degree of freedom. In particular, the crucial difference with the Heisenberg description is that the Hubbard model retains the electron-hole continuum, whose lower edge at $\omega=\Delta_c$ is below the onset of the two-magnon continuum: $\Delta_c < 2 \Delta_s$. We considered a half-filled bilayer, which includes a single ``$J_\text{eff}=1/2$'' effective orbital for each of the two Ir sites in the unit cell, following methods developed in parallel with this experimental study \cite{Suwa21}. The model contains an effective Coulomb repulsion $U$, and three electron hopping parameters: nearest and next-nearest in-plane hopping terms $t_\nu$ ($\nu = 1, 2$) within each Ir layer, and the spin-dependent hopping strength $t_z(\alpha)$ between Ir layers (Fig.~\ref{fig:ldep}b). $t_z(\alpha)$ is composed of an amplitude $|t_z|$ and a phase $\alpha$ arising from the appreciable \gls*{SOC} in the material (further details are given in the Methods section) \cite{Carter2013theory}. The model was solved using the \gls*{RPA} in the thermodynamic limit (\gls*{SI} Section 2), which is valid for intermediately correlated materials even at finite temperature \cite{Hirschmeier2015}. We constrain $t_\nu$ and $|t_z|$ to values compatible with density functional theory and photo-emission measurements and consider the effective $U$, which is strongly influenced by screening, as the primary tuning parameter \cite{Carter2013microscopic}. Figures \ref{fig:dispersion}c,d show the results of calculations with $t_1 = 0.115$~eV, $t_2 = 0.012$~eV, $|t_z| = 0.084$~eV, $\alpha = 1.41$, and $U = 0.325$~eV. The small $U$ is due to the extended Ir orbitals and because this effective parameterization reflects the difference between on-site and longer-range interactions in the real material.  The model identifies the quasiparticle dispersion at $q_c = 0$ as the transverse mode with a persistent well-defined nature even at high energies. Above the transverse mode, the spin response is fundamentally influenced by the finite charge gap. A broad continuum involving electron-hole spin transitions across the charge gap is present for all $q_c$ values covering a broad energy-momentum range. A new mode emerges around $(0, 0)$ and $(0.5, 0.5)$ for $q_c = 0.5$, which we identify as the excitonic longitudinal mode. 

To understand the excitonic longitudinal mode discussed, we first note that the tight-binding band structure analysis of Sr$_3$Ir$_2$O$_7$ suggests that it would be a narrow-gap band insulator or semi-metal even when Coulomb repulsion is neglected \cite{Carter2013microscopic}. This occurs due to bonding-antibonding band splitting arising from the bilayer hopping alongside \gls*{SOC}, generating a minimum of the conduction band dispersion near the Brillouin zone center and a maximum in the valence band dispersion near the antiferromagnetic zone center. A finite value of $U$ in a quasi-two-dimensional bilayer structure such as Sr$_3$Ir$_2$O$_7$ produces an attractive particle-hole interaction in the triplet channel because of the well-known direct-exchange mechanism. In turn, particle-hole pairs at wavevectors favored by the band structure form bound states, i.e.\ excitons, in the magnetic channel appearing at $q_c=0.5$, because of the odd parity of the exciton under exchange of the two layers. The spin anisotropy arising from \gls*{SOC} splits the exciton triplet into a low-energy state with $c$-axis spin quantum number $S^z=0$ and higher energy $S^z=\pm1$ states. Strictly speaking, \gls*{SOC} means that total spin is not a good quantum number, but we retain the singlet-triplet labels for clarity. As shown in the schematic representation in Fig.~\ref{fig:schematic}, the $S^z=0$ exciton condenses to form magnetic order at a wavevector of $(0.5, 0.5)$ ($q_c=0.5$). The  corresponding \gls*{QCP}, which exists at $U=U_\text{c}=0.27$~eV (for $t_1 = 0.115$~eV), then signals the  onset of the antiferromagnetic excitonic insulator state in Sr$_3$Ir$_2$O$_7$. Within the ordered state, what was a gapless $S^z=0$ exciton mode at $U=U_c$ becomes a gapped excitonic longitudinal mode for $U > U_c$. The existence and relatively low energy of this mode implies that $U$ in Sr$_3$Ir$_2$O$_7$ is only slightly above $U_c$. This property, together with the sufficiently large transverse mode gap $\Delta_s$, protects the excitonic longitudinal mode from decay into pairs of transverse modes. The longitudinal mode's bound electron-hole pair nature is especially vividly illustrated by its smooth merging with the particle-hole continuum away from $(0.5,0.5)$ and $(0,0)$. We plot the layer-resolved charge structure of the exciton in \gls*{SI} Section 3.

When heating an antiferromagnetic excitonic insulator, thermal fluctuations modify the magnetic properties via two different processes.  The first one corresponds to the destruction of N\'{e}el order via softening of the longitudinal mode. This softening signals the exciton condensation below $T=T_N$. The second process, that takes place at a higher temperature $T^*$, corresponds to thermal breaking of the excitons (unbinding of particle-hole pairs). A \gls*{RIXS} temperature series designed to test this idea at different high symmetry locations is plotted in Fig.~\ref{fig:tdep}a-d (linecuts at selected temperatures are shown in Fig.~S2). As expected, heating up from base temperature towards $T_\text{N}$ enhances the decay of the modes into the electron-hole continuum broadening the spectra and making it difficult to isolate the two modes in a single  spectrum.  We can, however, leverage the symmetry properties of the modes at different reciprocal space points to clarify the soft mode phenomenology. Since the transverse mode occurs at the same energy independent of $q_c$, and the longitudinal mode is present at $q_c = 0.5$ and absent at $q_c = 0$, the transverse mode temperature dependence can be studied in isolation at $q_c = 0$ (Fig.~\ref{fig:tdep}a,b). We observe that this mode has only minimal detectable softening, which is expected in view of the Ising nature of magnetism. In contrast, a substantial softening is seen at  $(0.5, 0.5)$ in Fig.~\ref{fig:tdep}d. Although both modes are present at $q_c = 0.5$, we know from $q_c = 0$ measurements that the transverse mode displays only minimal softening. Thus the longitudinal mode must play a major role in the softening to form the antiferromagnetic state. Our observed phenomenology is only captured with the intermediate coupling regime $(U/t_1=2.83)$ that we conclude is relevant for Sr$_3$Ir$_2$O$_7$. The strong coupling limit ($U/t_1 \gg 1$) would require a charge gap much larger than the observed values of 100-200~meV and, to our knowledge, it has not been able to predict any aspects of the temperature-dependent phenomenology of Sr$_3$Ir$_2$O$_7$. The excitonic insulator model is also supported by our temperature-dependent calculations, which are shown as dashed lines in Figs.~\ref{fig:tdep}c and d. Full calculations are shown in Fig.~S5 and explained in \gls*{SI} Section 2. Theory shows that exciton formation takes place at $T^* \approx 2T_\text{N}$, controlled by the exciton binding energy, which is of order the charge gap minus the longitudinal mode energy at the ordering wavevector. The mean-field transition temperature prediction is $T_\text{N}=424$~K, which is not too far above the measured $T_\text{N}=285$~K and which is expected since fluctuations are expected to reduce $T_\text{N}$ below the mean-field prediction. The predictions in Figs.~\ref{fig:tdep}c and d are shown with temperatures re-normalized to the experimental $T_\text{N}$.

The involvement of the longitudinal mode in magnetic long-range order is also evident from the temperature dependent quasi-elastic intensity. While most spectra feature the expected gradual enhancement in the quasi-elastic channel upon increasing temperature (Figs.~\ref{fig:tdep}e for (0, 0) and S2 for other reciprocal-lattice positions), the $(0.5, 0.5)$ spectrum at $q_c = 0.5$ displays a pronounced rise of intensity around $T_\text{N}$ (Figs.~\ref{fig:tdep}f). Note that neither $q_c = 0$ nor $q_c = 0.5$ correspond to the magnetic Bragg peak location, because the bilayer separation is incommensurate with respect to the $c$-axis lattice constant. Since in our setup $q_c = 0$ is closer to a magnetic Bragg peak than $q_c = 0.5$, we can exclude critical scattering from the long-range antiferromagnetic order as a significant contributor to this intensity as it would predict the opposite intensity behavior to what we observe (a more extensive demonstration of this is in \gls*{SI} Section 5). Thus the observed quasi-elastic anomaly at $T_\text{N}$ is indicative of substantial longitudinal mode condensation.  The excitonic insulator character of the ground state is further supported by a large increase in resistivity below $T_\text{N}$ (see see Fig.~\ref{fig:tdep}f) \cite{Cao2002anomalous}, as the condensation of the excitonic mode leads to a reduction in the electronic carriers participating in electrical transport. This property is distinct from what is expected for a strongly-coupled Mott insulator (i.e.\ the large $U$ limit of Fig.~\ref{fig:schematic}) where all charge-related processes are frozen out. The resistivity increase below $T_\text{N}$ could, in principle, also arise from Slater-type interactions, which can open a charge gap upon magnetic ordering. Sr$_3$Ir$_2$O$_7$, however, lacks strong Fermi surface nesting  \cite{Carter2013microscopic, Wang2013, King2013spectroscopic, Torre2014coherent} and is in the intermediately correlated ($t_1 \sim U$) rather than the weakly correlated ($t_1 \gg U$) regime, so the Slater mechanism is expected to have minimal relevance.

\section*{Discussion}

In summary, we have isolated and characterized a longitudinal magnetic mode in Sr$_3$Ir$_2$O$_7$, which merges with the electron-hole continuum at certain points in the Brillouin zone, and which softens upon heating concurrent with a decrease in the material's resistivity. These properties are consistent with those of an antiferromagnetic excitonic insulator state \cite{Halperin1968possible}. We substantiate this via calculations of a bilayer Hubbard model, in which electron-hole pairs are bound by magnetic exchange interactions between the electron and hole. This consistently explains all the electronic and magnetic properties of Sr$_3$Ir$_2$O$_7$ based on only one free parameter $U$, since all other parameter are strongly constrained by the electronic band structure of the material. The totality of these results identifies Sr$_3$Ir$_2$O$_7$ as a compelling candidate for the long-sought-after antiferromagnetic excitonic insulator.

Looking to the future, the intrinsically coupled spin and charge degrees of freedom in this state could have potential for realizing new functionalities \cite{Tokura2017emergent}, and suitably tuned material and/or laser-based approaches could realize methods to photo-excite these modes \cite{Kang2020coherent}. Further research on the topic may also include efforts to identify materials closer to the \gls*{QCP}, which in our study occurs at $U/t_1 = 2.35$. This could extend the reciprocal-space regions where the excitonic longitudinal mode exists. Another interesting direction would involve identifying excitonic easy-plane, rather than easy-axis, bilayer systems. These would host a different kind of soft excitonic longitudinal mode, often called ``Higgs'' mode, and could be used to study Higgs decay and renormalization effects in the presence of strong charge fluctuations. Careful selection of materials with multiple active orbitals could realize orbitally-ordered excitonic insulator states. Experimental realizations using chemical substitutions, strained thin films, high pressure or different bilayer materials, including ruthenates, osmates and other iridates, may help to answer some of these intriguing questions.

\section*{Methods\label{Methods}}
\noindent\textbf{Samples.}
Sr$_3$Ir$_2$O$_7$ single crystals were synthesized using the flux method \cite{Li2013}. Starting materials of IrO$_2$, SrCO$_3$, and SrCl$_2\cdot$6H$_2$O were mixed with a molar ratio of 2:1:20, and heated at 1200$^\circ$C for 10 hours in a platinum crucible. The melt was then cooled to 800$^\circ$C  at a rate of 3$^\circ$C/hour, before quenching to room temperature.  We index reciprocal space using a pseudo-tetragonal unit cell with $a = b = 3.896$~\AA{} and $c = 20.88$~\AA{} at room temperature.

 \noindent\textbf{Resonant inelastic X-ray scattering (RIXS) setup.}
\gls*{RIXS} spectra were measured at the 27-ID-B station of the Advanced Photon Source at Argonne National Laboratory. The incident x-ray beam was tuned to the Ir $L_3$-edge at 11.215~keV and monochromated using a Si (884) channel-cut monochromator. The exact x-ray energy was refined via a resonant energy of a standard IrO$_2$ and the Sr$_3$Ir$_2$O$_7$ sample and was set 3~eV below the resonant edge. Scattered photons were analyzed using a spherically bent diced silicon (844) analyzer with a curvature radius of 2~m. The energy and Q-resolution were 32.0(2)~meV and 0.105~\AA$^{-1}$ \gls*{FWHM}, respectively. A small background contribution arising from air scattering was removed by subtracting a constant value from the measured intensity. The value was determined by fitting the intensity on the energy-gain side of the spectra. 

The $L$ values in Fig.~\ref{fig:ldep}c-e were chosen such that they correspond to specific reciprocal-lattice positions with respect to the Ir-Ir interlayer spacing (see also Fig.~\ref{fig:ldep}a), i.e.\ $G + q_c = Ld/c$, where $G$ is an integer, $q_c$ the reduced $c$-axis reciprocal lattice position in terms of the Ir-Ir spacing, $d = 4.07$~\AA$^{}$ the shortest Ir-Ir interlayer spacing and $c = 20.88$~\AA{} the out-of-plane lattice constant. $q_c$ equals 0, 0.25 and 0.5 for $L = 25.65, 26.95$ and 28.25, respectively. 

The magnetic dispersions in Fig.~\ref{fig:dispersion}a,b were measured along $(H_1,K_1,25.65)$ and $(H_2,K_2,28.25)$ with $H_1$ and $K_1$ ranging between 0.5 and 1 and $H_2$ and $K_2$ between 0 and 0.5. The particular Brillouin zones were chosen to ensure a scattering geometry close to 90$^\circ$, minimizing Thompson scattering. For $(0,0,25.65)$, $(1,1,25.65)$, $(0,0,26.92)$ and $(0,0,28.25)$, $2\theta$ = 85.5, 90.2, 90.9 and 96.8$^\circ$, respectively. The sample was aligned in the horizontal $(H, H, L)$ scattering plane, such that the both dispersions could be probed through a sample rotation of $\Delta\chi\leq$ 4.1$^\circ$ relative to the surface normal.

\noindent\textbf{Analysis of the \gls*{RIXS} data.}
The spectra were analyzed by decomposing them into four components: (1) A quasi-elastic contribution (possibly containing contributions from phonons) which was modeled using  a pseudo-Voigt energy resolution function, along with an additional low-energy feature, which was modeled using the resolution functions at $\pm32$~meV, whose relative weights were constrained to follow the Bose factor. (2) The transverse magnetic mode was accounted for by a pseudo-Voigt function multiplied with an error function to capture the high-energy tail arising from the interactions with continuum. The interactions are enhanced when the modes and the continuum are less separated in energy, which leads to a reduced quasiparticle lifetime. In this case, we used a damped harmonic oscillator (with Bose factor) that was convoluted with the resolution function, which was further multiplied by an error function. (3) The longitudinal mode was described by either a pseudo-Voigt function or a damped harmonic oscillator, depending on whether or not it was resolution limited. (4) The magnetic continuum was reproduced using a broad damped harmonic oscillator multiplied by an error function to mimic its onset.

The excitonic longitudinal mode is strongly $q_c$ dependent, whereas the transverse magnetic mode and the magnetic continuum vary very weakly with $q_c$. Thus, we analyzed the spectra measured at $q_c = 0$ and $q_c = 0.5$ simultaneously to disentangle the excitonic contribution from the other components. The positions and lineshapes of the transverse magnetic mode and the electron-hole magnetic continuum were constrained to be independent of $q_c$, i.e.\ only the amplitudes were varied. The extra peaks at $q_c = 0.5$ give information about the excitonic longitudinal mode. During the procedure, the elastic energy was allowed to vary to correct for small fluctuations of the incident energy.

\noindent\textbf{Theoretical Model.}
Sr$_3$Ir$_2$O$_7$ hosts Ir$^{4+}$ ions, which have 5 electrons in the active Ir $5d^5$ valance band. The dominant splitting of this band comes from the close-to-cubic crystal field leaving empty $e_g$ states and 5 electrons in the $t_{2g}$ states. \gls*{SOC} further splits the $t_{2g}$ manifold into a full $J_\text{eff}=3/2$ orbital a half-filled $J_\text{eff}=1/2$ orbital at the Fermi level \cite{Kim2008novel}. Our model involves projecting the band structure onto this $J_\text{eff}=1/2$ doublet. The basic structural unit, shown in Fig.~\ref{fig:ldep}b, contains two Ir atoms, so the experimental data were interpreted using a half-filled bilayer Hubbard model $H=-H_{\rm K}+H_{\rm I}$ with $H_{\rm I}=U\sum_{\bm r} n_{\bm{r}\uparrow} n_{\bm{r}\downarrow}$ and
\begin{equation}
\begin{split}
H_{\rm K} = \sum_{\bm{r},\bm{\delta}_\nu}t_\nu c_{\bm{r}}^\dagger c_{\bm{r}+\bm{\delta}_\nu}
+\sum_{\bm{r}_\bot}c_{(\bm{r}_\bot,1)}^\dagger t_z(\alpha) c_{(\bm{r}_\bot,2)} + {\rm H.c}. ,
	\label{Halt}
\end{split}
\end{equation}
where $t_\nu$ ($\nu$ = 1, 2) are the nearest- and next-nearest-neighbour hopping amplitudes within the square lattice of each Ir-layer, and $t_z(\alpha) = |t_z| e^{i\frac{\alpha}{2}\varepsilon_{\bm{r}}\sigma_z}$, with $\sigma_z$ the Pauli matrix  describes the $J_\text{eff}$ spin dependent hopping strength between layers. The overall phase was chosen to gauge away the phase for $t_\nu$. The operator $ c_{\bm{r}}^{\dagger}$ = [$ c_{\uparrow,\bm{r}}^{\dagger}$, $c_{\downarrow,\bm{r}}^{\dagger}$] creates the Nambu spinor of the electron field at $\bm{r} = (\bm{r}_\bot, l)$ with $l = 1, 2$ denoting the layer index and $\bm{r}_\bot = r_1\bm{a}_1 + r_2\bm{a}_2$. Here, the primitive in-plane lattice vectors are denoted by $\bm{a}_1$ and $\bm{a}_2$, and the directed neighbouring bonds are represented by $\bm{\delta}_1=\bm{a}_1,\bm{a}_2$ and $\bm{\delta}_2=\bm{a}_1 \pm \bm{a}_2$. In the interaction term $H_{\rm I}$, $U$ is the effective Coulomb interaction, and $n_{\bm{r}\sigma}$ is the density operator for electrons of spin $\sigma$ at $\bm{r}$. In the spin-dependent hopping term, the sign $\varepsilon_{\bm{r}}$ takes the values $\pm1$ depending on which sublattice of the bipartite bilayer system $\bm{r}$ points to. The phase $\alpha$ arises from hopping matrix elements between $d_{xz}$ and $d_{yz}$ orbitals, which are allowed through the staggered octahedral rotations in the unit cell along side \gls*{SOC} \cite{Carter2013theory, Cao2018}. In the model, \gls*{SOC} enters via the phase of the $c$-axis hopping, which is smaller than the in-plane bandwidth, justifying the approximate use of singlet and triplet for labels of the different excitons. The model was studied at half-filling in the sense that it contains two bands (bonding and anti-bonding) in the model, which host two electrons as is appropriate for Sr$_3$Ir$_2$O$_7$ \cite{Carter2013theory, Wang2013, King2013spectroscopic, Torre2014coherent}. We solved the model using the \gls*{RPA} in the thermodynamic limit (detailed information is given in \gls*{SI} Section 2), which is valid for intermediately correlated materials even at finite temperatures \cite{Hirschmeier2015}. The theoretically determined N\'{e}el temperature in this case is $T^{\text{cal}}_{\text{N}} = 424$~K which is slightly larger than the experimental value $T_{\text{N}} = 285$~K. This is expected within the \gls*{RPA} we use here, as this ignores fluctuations than act to reduce the transition temperature. The dynamical spin structure factors in Figs.~\ref{fig:dispersion}c and d are shown after convolution with the experimental resolution.

A more complex model could include all $t_{2g}$ or all $d$ orbitals, rather than just effective $J_\text{eff}=1/2$ doublets. The success of our $J_\text{eff}=1/2$ only model suggests that orbital degrees of freedom are entirely frozen-out of the problem or manifest themselves in very subtle ways beyond current detection limits. Due to this, the Sr$_3$Ir$_2$O$_7$ excitonic insulator state has no orbital component (other than in the trivial sense that the  $J_\text{eff}=1/2$ states in themselves are a coupled modulation of spin and orbital angular momentum). A possible \gls*{SOC}-induced orbital order is discussed in \gls*{SI} Section 6.

\section*{References}
\bibliographystyle{naturemag}
\bibliography{refs}

\begin{thebibliography}{10}
\expandafter\ifx\csname url\endcsname\relax
  \def\url#1{\texttt{#1}}\fi
\expandafter\ifx\csname urlprefix\endcsname\relax\def\urlprefix{URL }\fi
\providecommand{\bibinfo}[2]{#2}
\providecommand{\eprint}[2][]{\url{#2}}

\bibitem{Mott1961transition}
\bibinfo{author}{Mott, N.~F.}
\newblock \bibinfo{title}{The transition to the metallic state}.
\newblock \emph{\bibinfo{journal}{Philos. Mag.}} \textbf{\bibinfo{volume}{6}},
  \bibinfo{pages}{287--309} (\bibinfo{year}{1961}).

\bibitem{Keldysh1965possible}
\bibinfo{author}{Keldysh, L.} \& \bibinfo{author}{Kopaev, Y.~V.}
\newblock \bibinfo{title}{Possible instability of semimetallic state toward
  {Coulomb} interaction}.
\newblock \emph{\bibinfo{journal}{Soviet Physics Solid State, USSR}}
  \textbf{\bibinfo{volume}{6}}, \bibinfo{pages}{2219–2224}
  (\bibinfo{year}{1965}).

\bibitem{Jerome1967excitonic}
\bibinfo{author}{J\'erome, D.}, \bibinfo{author}{Rice, T.~M.} \&
  \bibinfo{author}{Kohn, W.}
\newblock \bibinfo{title}{Excitonic insulator}.
\newblock \emph{\bibinfo{journal}{Phys. Rev.}} \textbf{\bibinfo{volume}{158}},
  \bibinfo{pages}{462--475} (\bibinfo{year}{1967}).

\bibitem{Halperin1968possible}
\bibinfo{author}{Halperin, B.~I.} \& \bibinfo{author}{Rice, T.~M.}
\newblock \bibinfo{title}{Possible anomalies at a semimetal-semiconductor
  transistion}.
\newblock \emph{\bibinfo{journal}{Rev. Mod. Phys.}}
  \textbf{\bibinfo{volume}{40}}, \bibinfo{pages}{755--766}
  (\bibinfo{year}{1968}).

\bibitem{Bucher1991excitonic}
\bibinfo{author}{Bucher, B.}, \bibinfo{author}{Steiner, P.} \&
  \bibinfo{author}{Wachter, P.}
\newblock \bibinfo{title}{Excitonic insulator phase in
  {${\mathrm{TmSe}}_{0.45}{\mathrm{Te}}_{0.55}$}}.
\newblock \emph{\bibinfo{journal}{Phys. Rev. Lett.}}
  \textbf{\bibinfo{volume}{67}}, \bibinfo{pages}{2717--2720}
  (\bibinfo{year}{1991}).

\bibitem{Wakisaka2009excitonic}
\bibinfo{author}{Wakisaka, Y.} \emph{et~al.}
\newblock \bibinfo{title}{Excitonic insulator state in
  {${\mathrm{Ta}}_{2}{\mathrm{NiSe}}_{5}$} probed by photoemission
  spectroscopy}.
\newblock \emph{\bibinfo{journal}{Phys. Rev. Lett.}}
  \textbf{\bibinfo{volume}{103}}, \bibinfo{pages}{026402}
  (\bibinfo{year}{2009}).

\bibitem{Eisenstein2014exciton}
\bibinfo{author}{Eisenstein, J.}
\newblock \bibinfo{title}{Exciton condensation in bilayer quantum hall
  systems}.
\newblock \emph{\bibinfo{journal}{Annu. Rev. Condens. Matter Phys.}}
  \textbf{\bibinfo{volume}{5}}, \bibinfo{pages}{159--181}
  (\bibinfo{year}{2014}).

\bibitem{Kogar2017signatures}
\bibinfo{author}{Kogar, A.} \emph{et~al.}
\newblock \bibinfo{title}{Signatures of exciton condensation in a transition
  metal dichalcogenide}.
\newblock \emph{\bibinfo{journal}{Science}} \textbf{\bibinfo{volume}{358}},
  \bibinfo{pages}{1314--1317} (\bibinfo{year}{2017}).

\bibitem{Lu2017zero}
\bibinfo{author}{Lu, Y.} \emph{et~al.}
\newblock \bibinfo{title}{Zero-gap semiconductor to excitonic insulator
  transition in {Ta$_2$NiSe$_5$}}.
\newblock \emph{\bibinfo{journal}{Nature communications}}
  \textbf{\bibinfo{volume}{8}}, \bibinfo{pages}{1--7} (\bibinfo{year}{2017}).

\bibitem{Slater1951magnetic}
\bibinfo{author}{Slater, J.~C.}
\newblock \bibinfo{title}{Magnetic effects and the hartree-fock equation}.
\newblock \emph{\bibinfo{journal}{Phys. Rev.}} \textbf{\bibinfo{volume}{82}},
  \bibinfo{pages}{538--541} (\bibinfo{year}{1951}).

\bibitem{Moon2008}
\bibinfo{author}{Moon, S.~J.} \emph{et~al.}
\newblock \bibinfo{title}{Dimensionality-controlled insulator-metal transition
  and correlated metallic state in $5d$ transition metal oxides
  {Sr$_{n+1}$Ir$_n$O$_{3n+1}$}: ($n=1$, 2, and $\ensuremath{\infty}$)}.
\newblock \emph{\bibinfo{journal}{Phys. Rev. Lett.}}
  \textbf{\bibinfo{volume}{101}}, \bibinfo{pages}{226402}
  (\bibinfo{year}{2008}).

\bibitem{Kim2012_2}
\bibinfo{author}{Kim, J.} \emph{et~al.}
\newblock \bibinfo{title}{Large spin-wave energy gap in the bilayer iridate
  {Sr$_3$Ir$_2$O$_7$}: Evidence for enhanced dipolar interactions near the
  {Mott} metal-insulator transition}.
\newblock \emph{\bibinfo{journal}{Phys. Rev. Lett.}}
  \textbf{\bibinfo{volume}{109}}, \bibinfo{pages}{157402}
  (\bibinfo{year}{2012}).

\bibitem{Lohofer2015dynamical}
\bibinfo{author}{Loh\"ofer, M.} \emph{et~al.}
\newblock \bibinfo{title}{Dynamical structure factors and excitation modes of
  the bilayer {Heisenberg} model}.
\newblock \emph{\bibinfo{journal}{Phys. Rev. B}} \textbf{\bibinfo{volume}{92}},
  \bibinfo{pages}{245137} (\bibinfo{year}{2015}).

\bibitem{Moretti2015evidence}
\bibinfo{author}{Moretti~Sala, M.} \emph{et~al.}
\newblock \bibinfo{title}{Evidence of quantum dimer excitations in
  {${\mathrm{Sr}}_{3}{\mathrm{Ir}}_{2}{\mathrm{O}}_{7}$}}.
\newblock \emph{\bibinfo{journal}{Phys. Rev. B}} \textbf{\bibinfo{volume}{92}},
  \bibinfo{pages}{024405} (\bibinfo{year}{2015}).

\bibitem{Zhou2020amplitude}
\bibinfo{author}{Zhou, C.}, \bibinfo{author}{Yan, Z.}, \bibinfo{author}{Sun,
  K.}, \bibinfo{author}{Starykh, O.~A.} \& \bibinfo{author}{Meng, Z.~Y.}
\newblock \bibinfo{title}{Amplitude mode in quantum magnets via dimensional
  crossover}.
\newblock \emph{\bibinfo{journal}{arXiv preprint arXiv:2007.12715}}
  (\bibinfo{year}{2020}).

\bibitem{Su2020stable}
\bibinfo{author}{Su, Y.} \emph{et~al.}
\newblock \bibinfo{title}{Stable {Higgs} mode in anisotropic quantum magnets}.
\newblock \emph{\bibinfo{journal}{Phys. Rev. B}}
  \textbf{\bibinfo{volume}{102}}, \bibinfo{pages}{125102}
  (\bibinfo{year}{2020}).

\bibitem{Hogan2016disordered}
\bibinfo{author}{Hogan, T.} \emph{et~al.}
\newblock \bibinfo{title}{Disordered dimer state in electron-doped
  {${\mathrm{Sr}}_{3}{\mathrm{Ir}}_{2}{\mathrm{O}}_{7}$}}.
\newblock \emph{\bibinfo{journal}{Phys. Rev. B}} \textbf{\bibinfo{volume}{94}},
  \bibinfo{pages}{100401} (\bibinfo{year}{2016}).

\bibitem{Gretarsson2016two}
\bibinfo{author}{Gretarsson, H.} \emph{et~al.}
\newblock \bibinfo{title}{Two-magnon raman scattering and pseudospin-lattice
  interactions in {${\mathrm{Sr}}_{2}{\mathrm{IrO}}_{4}$} and
  {${\mathrm{Sr}}_{3}{\mathrm{Ir}}_{2}{\mathrm{O}}_{7}$}}.
\newblock \emph{\bibinfo{journal}{Phys. Rev. Lett.}}
  \textbf{\bibinfo{volume}{116}}, \bibinfo{pages}{136401}
  (\bibinfo{year}{2016}).

\bibitem{Lu2017doping}
\bibinfo{author}{Lu, X.} \emph{et~al.}
\newblock \bibinfo{title}{Doping evolution of magnetic order and magnetic
  excitations in
  {$({\mathrm{Sr}}_{1\ensuremath{-}x}{\mathrm{La}}_{x}{)}_{3}{\mathrm{Ir}}_{2}{\mathrm{O}}_{7}$}}.
\newblock \emph{\bibinfo{journal}{Phys. Rev. Lett.}}
  \textbf{\bibinfo{volume}{118}}, \bibinfo{pages}{027202}
  (\bibinfo{year}{2017}).

\bibitem{Li2020symmetry}
\bibinfo{author}{Li, S.} \emph{et~al.}
\newblock \bibinfo{title}{Symmetry-resolved two-magnon excitations in a strong
  spin-orbit-coupled bilayer antiferromagnet}.
\newblock \emph{\bibinfo{journal}{Phys. Rev. Lett.}}
  \textbf{\bibinfo{volume}{125}}, \bibinfo{pages}{087202}
  (\bibinfo{year}{2020}).

\bibitem{Mohapatra2017}
\bibinfo{author}{Mohapatra, S.}, \bibinfo{author}{van~den Brink, J.} \&
  \bibinfo{author}{Singh, A.}
\newblock \bibinfo{title}{Magnetic excitations in a three-orbital model for the
  strongly spin-orbit coupled iridates: Effect of mixing between the
  $j=\frac{1}{2}$ and $\frac{3}{2}$ sectors}.
\newblock \emph{\bibinfo{journal}{Phys. Rev. B}} \textbf{\bibinfo{volume}{95}},
  \bibinfo{pages}{094435} (\bibinfo{year}{2017}).

\bibitem{Cao2002anomalous}
\bibinfo{author}{Cao, G.} \emph{et~al.}
\newblock \bibinfo{title}{Anomalous magnetic and transport behavior in the
  magnetic insulator {${\mathrm{Sr}}_{3}{\mathrm{Ir}}_{2}{\mathrm{O}}_{7}$}}.
\newblock \emph{\bibinfo{journal}{Phys. Rev. B}} \textbf{\bibinfo{volume}{66}},
  \bibinfo{pages}{214412} (\bibinfo{year}{2002}).

\bibitem{Okada2013imaging}
\bibinfo{author}{Okada, Y.} \emph{et~al.}
\newblock \bibinfo{title}{Imaging the evolution of metallic states in a
  correlated iridate}.
\newblock \emph{\bibinfo{journal}{Nature materials}}
  \textbf{\bibinfo{volume}{12}}, \bibinfo{pages}{707--713}
  (\bibinfo{year}{2013}).

\bibitem{Wang2013}
\bibinfo{author}{Wang, Q.} \emph{et~al.}
\newblock \bibinfo{title}{Dimensionality-controlled mott transition and
  correlation effects in single-layer and bilayer perovskite iridates}.
\newblock \emph{\bibinfo{journal}{Phys. Rev. B}} \textbf{\bibinfo{volume}{87}},
  \bibinfo{pages}{245109} (\bibinfo{year}{2013}).

\bibitem{King2013spectroscopic}
\bibinfo{author}{King, P. D.~C.} \emph{et~al.}
\newblock \bibinfo{title}{Spectroscopic indications of polaronic behavior of
  the strong spin-orbit insulator sr${}_{3}$ir${}_{2}$o${}_{7}$}.
\newblock \emph{\bibinfo{journal}{Phys. Rev. B}} \textbf{\bibinfo{volume}{87}},
  \bibinfo{pages}{241106} (\bibinfo{year}{2013}).

\bibitem{Torre2014coherent}
\bibinfo{author}{de~la Torre, A.} \emph{et~al.}
\newblock \bibinfo{title}{Coherent quasiparticles with a small fermi surface in
  lightly doped {${\mathrm{Sr}}_{3}{\mathrm{Ir}}_{2}{\mathrm{O}}_{7}$}}.
\newblock \emph{\bibinfo{journal}{Phys. Rev. Lett.}}
  \textbf{\bibinfo{volume}{113}}, \bibinfo{pages}{256402}
  (\bibinfo{year}{2014}).

\bibitem{Suwa21}
\bibinfo{author}{Suwa, H.}, \bibinfo{author}{Zhang, S.-S.} \&
  \bibinfo{author}{Batista, C.~D.}
\newblock \bibinfo{title}{Exciton condensation in bilayer spin-orbit
  insulator}.
\newblock \emph{\bibinfo{journal}{Phys. Rev. Research}}
  \textbf{\bibinfo{volume}{3}}, \bibinfo{pages}{013224} (\bibinfo{year}{2021}).

\bibitem{Carter2013theory}
\bibinfo{author}{Carter, J.-M.}, \bibinfo{author}{Shankar~V., V.} \&
  \bibinfo{author}{Kee, H.-Y.}
\newblock \bibinfo{title}{Theory of metal-insulator transition in the family of
  perovskite iridium oxides}.
\newblock \emph{\bibinfo{journal}{Phys. Rev. B}} \textbf{\bibinfo{volume}{88}},
  \bibinfo{pages}{035111} (\bibinfo{year}{2013}).

\bibitem{Hirschmeier2015}
\bibinfo{author}{Hirschmeier, D.}, \bibinfo{author}{Hafermann, H.},
  \bibinfo{author}{Gull, E.}, \bibinfo{author}{Lichtenstein, A.~I.} \&
  \bibinfo{author}{Antipov, A.~E.}
\newblock \bibinfo{title}{Mechanisms of finite-temperature magnetism in the
  three-dimensional {Hubbard} model}.
\newblock \emph{\bibinfo{journal}{Phys. Rev. B}} \textbf{\bibinfo{volume}{92}},
  \bibinfo{pages}{144409} (\bibinfo{year}{2015}).

\bibitem{Carter2013microscopic}
\bibinfo{author}{Carter, J.-M.} \& \bibinfo{author}{Kee, H.-Y.}
\newblock \bibinfo{title}{Microscopic theory of magnetism in
  {Sr${}_{3}$Ir${}_{2}$O${}_{7}$}}.
\newblock \emph{\bibinfo{journal}{Phys. Rev. B}} \textbf{\bibinfo{volume}{87}},
  \bibinfo{pages}{014433} (\bibinfo{year}{2013}).

\bibitem{Tokura2017emergent}
\bibinfo{author}{Tokura, Y.}, \bibinfo{author}{Kawasaki, M.} \&
  \bibinfo{author}{Nagaosa, N.}
\newblock \bibinfo{title}{Emergent functions of quantum materials}.
\newblock \emph{\bibinfo{journal}{Nature Physics}}
  \textbf{\bibinfo{volume}{13}}, \bibinfo{pages}{1056--1068}
  (\bibinfo{year}{2017}).

\bibitem{Kang2020coherent}
\bibinfo{author}{Kang, S.} \emph{et~al.}
\newblock \bibinfo{title}{Coherent many-body exciton in van der {Waals}
  antiferromagnet {NiPS$_3$}}.
\newblock \emph{\bibinfo{journal}{Nature}} \textbf{\bibinfo{volume}{583}},
  \bibinfo{pages}{785--789} (\bibinfo{year}{2020}).

\bibitem{Li2013}
\bibinfo{author}{Li, L.} \emph{et~al.}
\newblock \bibinfo{title}{Tuning the ${J}_{\mathrm{eff}}=\frac{1}{2}$
  insulating state via electron doping and pressure in the double-layered
  iridate {Sr${}_{3}$Ir${}_{2}$O${}_{7}$}}.
\newblock \emph{\bibinfo{journal}{Phys. Rev. B}} \textbf{\bibinfo{volume}{87}},
  \bibinfo{pages}{235127} (\bibinfo{year}{2013}).

\bibitem{Kim2008novel}
\bibinfo{author}{Kim, B.~J.} \emph{et~al.}
\newblock \bibinfo{title}{Novel {${J}_{\mathrm{eff}}=1/2$ Mott} state induced
  by relativistic spin-orbit coupling in
  {${\mathrm{Sr}}_{2}{\mathrm{IrO}}_{4}$}}.
\newblock \emph{\bibinfo{journal}{Phys. Rev. Lett.}}
  \textbf{\bibinfo{volume}{101}}, \bibinfo{pages}{076402}
  (\bibinfo{year}{2008}).

\bibitem{Cao2018}
\bibinfo{author}{Cao, G.} \& \bibinfo{author}{Schlottmann, P.}
\newblock \bibinfo{title}{The challenge of spin{\textendash}orbit-tuned ground
  states in iridates: a key issues review}.
\newblock \emph{\bibinfo{journal}{Reports on Progress in Physics}}
  \textbf{\bibinfo{volume}{81}}, \bibinfo{pages}{042502}
  (\bibinfo{year}{2018}).

\bibitem{Mazzone2021data}
\bibinfo{author}{Mazzone, D.~G.} \emph{et~al.}
\newblock \bibinfo{title}{{Data Repository for: Antiferromagnetic Excitonic
  Insulator State in Sr$_3$Ir$_2$O$_7$}} (\bibinfo{year}{2022}).
\newblock \urlprefix\url{https://doi.org/10.5281/zenodo.5812989}.

\end{thebibliography}


\begin{thebibliography}{10}
\expandafter\ifx\csname url\endcsname\relax
  \def\url#1{\texttt{#1}}\fi
\expandafter\ifx\csname urlprefix\endcsname\relax\def\urlprefix{URL }\fi
\providecommand{\bibinfo}[2]{#2}
\providecommand{\eprint}[2][]{\url{#2}}

\bibitem{Kim2012_2}
\bibinfo{author}{Kim, J.} \emph{et~al.}
\newblock \bibinfo{title}{Large spin-wave energy gap in the bilayer iridate
  {Sr$_3$Ir$_2$O$_7$}: Evidence for enhanced dipolar interactions near the
  {Mott} metal-insulator transition}.
\newblock \emph{\bibinfo{journal}{Phys. Rev. Lett.}}
  \textbf{\bibinfo{volume}{109}}, \bibinfo{pages}{157402}
  (\bibinfo{year}{2012}).

\bibitem{Moretti2015evidence}
\bibinfo{author}{Moretti~Sala, M.} \emph{et~al.}
\newblock \bibinfo{title}{Evidence of quantum dimer excitations in
  {${\mathrm{Sr}}_{3}{\mathrm{Ir}}_{2}{\mathrm{O}}_{7}$}}.
\newblock \emph{\bibinfo{journal}{Phys. Rev. B}} \textbf{\bibinfo{volume}{92}},
  \bibinfo{pages}{024405} (\bibinfo{year}{2015}).

\bibitem{Gretarsson2016two}
\bibinfo{author}{Gretarsson, H.} \emph{et~al.}
\newblock \bibinfo{title}{Two-magnon raman scattering and pseudospin-lattice
  interactions in {${\mathrm{Sr}}_{2}{\mathrm{IrO}}_{4}$} and
  {${\mathrm{Sr}}_{3}{\mathrm{Ir}}_{2}{\mathrm{O}}_{7}$}}.
\newblock \emph{\bibinfo{journal}{Phys. Rev. Lett.}}
  \textbf{\bibinfo{volume}{116}}, \bibinfo{pages}{136401}
  (\bibinfo{year}{2016}).

\bibitem{Hogan2016disordered}
\bibinfo{author}{Hogan, T.} \emph{et~al.}
\newblock \bibinfo{title}{Disordered dimer state in electron-doped
  {${\mathrm{Sr}}_{3}{\mathrm{Ir}}_{2}{\mathrm{O}}_{7}$}}.
\newblock \emph{\bibinfo{journal}{Phys. Rev. B}} \textbf{\bibinfo{volume}{94}},
  \bibinfo{pages}{100401} (\bibinfo{year}{2016}).

\bibitem{Li2020symmetry}
\bibinfo{author}{Li, S.} \emph{et~al.}
\newblock \bibinfo{title}{Symmetry-resolved two-magnon excitations in a strong
  spin-orbit-coupled bilayer antiferromagnet}.
\newblock \emph{\bibinfo{journal}{Phys. Rev. Lett.}}
  \textbf{\bibinfo{volume}{125}}, \bibinfo{pages}{087202}
  (\bibinfo{year}{2020}).

\bibitem{Carter2013theory}
\bibinfo{author}{Carter, J.-M.}, \bibinfo{author}{Shankar~V., V.} \&
  \bibinfo{author}{Kee, H.-Y.}
\newblock \bibinfo{title}{Theory of metal-insulator transition in the family of
  perovskite iridium oxides}.
\newblock \emph{\bibinfo{journal}{Phys. Rev. B}} \textbf{\bibinfo{volume}{88}},
  \bibinfo{pages}{035111} (\bibinfo{year}{2013}).

\bibitem{Igarashi2014analysis}
\bibinfo{author}{Igarashi, J.-i.} \& \bibinfo{author}{Nagao, T.}
\newblock \bibinfo{title}{Analysis of resonant inelastic x-ray scattering from
  {${\mathrm{Sr}}_{2}{\mathrm{IrO}}_{4}$} in an itinerant-electron approach}.
\newblock \emph{\bibinfo{journal}{Phys. Rev. B}} \textbf{\bibinfo{volume}{90}},
  \bibinfo{pages}{064402} (\bibinfo{year}{2014}).

\bibitem{Fujita2021magnetic}
\bibinfo{author}{Fujita, M.}, \bibinfo{author}{Ikeuchi, K.},
  \bibinfo{author}{Kajimoto, R.} \& \bibinfo{author}{Nakamura, M.}
\newblock \bibinfo{title}{Magnetic excitations of {Sr$_3$Ir$_2$O$_7$} observed
  by inelastic neutron scattering technique}.
\newblock \emph{\bibinfo{journal}{Journal of the Physical Society of Japan}}
  \textbf{\bibinfo{volume}{90}}, \bibinfo{pages}{025001}
  (\bibinfo{year}{2021}).
\newblock \eprint{https://doi.org/10.7566/JPSJ.90.025001}.

\bibitem{Suwa21}
\bibinfo{author}{Suwa, H.}, \bibinfo{author}{Zhang, S.-S.} \&
  \bibinfo{author}{Batista, C.~D.}
\newblock \bibinfo{title}{Exciton condensation in bilayer spin-orbit
  insulator}.
\newblock \emph{\bibinfo{journal}{Phys. Rev. Research}}
  \textbf{\bibinfo{volume}{3}}, \bibinfo{pages}{013224} (\bibinfo{year}{2021}).

\bibitem{Khaliullin2013excitonic}
\bibinfo{author}{Khaliullin, G.}
\newblock \bibinfo{title}{Excitonic magnetism in van vleck--type ${d}^{4}$ mott
  insulators}.
\newblock \emph{\bibinfo{journal}{Phys. Rev. Lett.}}
  \textbf{\bibinfo{volume}{111}}, \bibinfo{pages}{197201}
  (\bibinfo{year}{2013}).

\bibitem{Jain2017higgs}
\bibinfo{author}{Jain, A.} \emph{et~al.}
\newblock \bibinfo{title}{Higgs mode and its decay in a two-dimensional
  antiferromagnet}.
\newblock \emph{\bibinfo{journal}{Nature Physics}}
  \textbf{\bibinfo{volume}{13}}, \bibinfo{pages}{633--637}
  (\bibinfo{year}{2017}).

\bibitem{Halperin1968possible}
\bibinfo{author}{Halperin, B.~I.} \& \bibinfo{author}{Rice, T.~M.}
\newblock \bibinfo{title}{Possible anomalies at a semimetal-semiconductor
  transistion}.
\newblock \emph{\bibinfo{journal}{Rev. Mod. Phys.}}
  \textbf{\bibinfo{volume}{40}}, \bibinfo{pages}{755--766}
  (\bibinfo{year}{1968}).

\end{thebibliography}

\section*{Acknowledgments}
We thank D.F.\ McMorrow, B.\ Normand, Ch.\ R\"{u}egg, and J.P.\ Hill for fruitful discussions. Work performed at Brookhaven National Laboratory was supported by the US Department of Energy, Division of Materials Science, under Contract No.~DESC0012704. This research used resources of the Advanced Photon Source, a U.S. Department of Energy (DOE) Office of Science User Facility operated for the DOE Office of Science by Argonne National Laboratory under Contract No. DE-AC02-06CH11357. This research used resources of the Oak Ridge Leadership Computing Facility at the Oak Ridge National Laboratory, which is supported by the Office of Science of the U.S. Department of Energy under Contract No. DE-AC05-00OR22725. D.G.M. acknowledges support from the Swiss National Science Foundation, Fellowship No. P2EZP2\_175092. H.\nobreak\,S. acknowledges support from JSPS KAKENHI Grant No.~JP19K14650. X.L.\ acknowledges the support from the National Natural Science Foundation of China under grant No.~11934017. K.J.\ and Y.G.S.\ acknowledge the support from the National Natural Science Foundation of China (Grants No.~U2032204), and the K.C.\ Wong Education Foundation (GJTD-2018-01). Y.G.S. acknowledges the Chinese National Key Research and Development Program (No.\ 2017YFA0302901) and the Strategic Priority Research Program (B) of the Chinese Academy of Sciences (Grant No.\ XDB33000000).  J.L.\ acknowledges support from the National Science Foundation under Grant No.~DMR-1848269. J.Y.\ acknowledges funding from the State of Tennessee and Tennessee Higher Education Commission (THEC) through their support of the Center for Materials Processing. H.M.\ was sponsored by the Laboratory Directed Research and Development Program of Oak Ridge National Laboratory, managed by UT-Battelle, LLC, for the U.S.\ Department of Energy.

\section*{Author contributions}
The project was conceived and initiated by M.P.M.D, C.D.B., J.L.\ and X.L. D.G.M., Y.S., G.F., J.Y., H.M., M.H.U., and D.M.C. prepared and performed the experiments. K.J. and Y.G.S. synthesised the samples. D.G.M.\ and Y.S.\ analyzed the experimental data, H.S. and S.S.Z.\ provided the theoretical calculations. The results were interpreted by D.G.M, Y.S, H.S., J.S., J.L. C.D.B, and M.P.M.D, The paper was written by D.G.M., Y.S., H.S., X.L., C.D.B., and M.P.M.D. with the input from all co-authors.

\section*{Additional Information}
Correspondence and requests for materials should be addressed to D.G.M, X.L. or M.P.M.D.
\section*{Competing financial interests}
The authors declare no competing interests.
\section*{Data and code availability}
The \gls*{RIXS} data generated in this study have been deposited in the Zenodo database under accession code 5812989 \cite{Mazzone2021data}. The code used in this study is available from the authors upon reasonable request.

\end{document}


\title{Antiferromagnetic Excitonic Insulator State in Sr$_3$Ir$_2$O$_7$\\
Supplementary Information}

\author{D. G. Mazzone}
\email{daniel.mazzone@psi.ch}
\affiliation{Department of Condensed Matter Physics and Materials Science, Brookhaven National Laboratory, Upton, New York 11973, USA}
\affiliation{Laboratory for Neutron Scattering and Imaging, Paul Scherrer Institut, CH-5232 Villigen, Switzerland}

\author{Y. Shen}
\affiliation{Department of Condensed Matter Physics and Materials Science, Brookhaven National Laboratory, Upton, New York 11973, USA}

\author{H. Suwa}
\affiliation{Department of Physics, The University of Tokyo, Tokyo 113-0033, Japan}
\affiliation{Department of Physics and Astronomy, University of Tennessee, Knoxville, Tennessee 37996, USA}

\author{G. Fabbris}
\affiliation{Advanced Photon Source, Argonne National Laboratory, Argonne, Illinois 60439, USA}

\author{J. Yang}
\affiliation{Department of Physics and Astronomy, University of Tennessee, Knoxville, Tennessee 37996, USA}

\author{S-S. Zhang}
\affiliation{Department of Physics and Astronomy, University of Tennessee, Knoxville, Tennessee 37996, USA}

\author{H. Miao}
\affiliation{Material Science and Technology Division, Oak Ridge National Laboratory, Oak Ridge, Tennessee 37831, USA}

\author{J. Sears}
\affiliation{Department of Condensed Matter Physics and Materials Science, Brookhaven National Laboratory, Upton, New York 11973, USA}

\author{Ke Jia}
\author{Y. Shi}
\affiliation{Beijing National Laboratory for Condensed Matter Physics, Institute of Physics, Chinese Academy of Sciences, Beijing 100190, China}

\author{M .H. Upton}
\author{D. M. Casa}
\affiliation{Advanced Photon Source, Argonne National Laboratory, Argonne, Illinois 60439, USA}

\author{X. Liu}
\email{liuxr@shanghaitec.edu.cn}
\affiliation{School of Physical Science and Technology, ShanghaiTech University, Shanghai 201210, China}

\author{J. Liu}
\affiliation{Department of Physics and Astronomy, University of Tennessee, Knoxville, Tennessee 37996, USA}

\author{C. Batista}
\affiliation{Department of Physics and Astronomy, University of Tennessee, Knoxville, Tennessee 37996, USA}
\affiliation{Quantum Condensed Matter Division and Shull-Wollan Center, Oak Ridge National Laboratory, Oak Ridge, TN 37831, USA}

\author{M. P. M. Dean}
\email{mdean@bnl.gov}
\affiliation{Department of Condensed Matter Physics and Materials Science, Brookhaven National Laboratory, Upton, New York 11973, USA}

\renewcommand{\thefigure}{S\arabic{figure}}

\newcommand\tlc[1]{\texorpdfstring{\lowercase{#1}}{#1}}


\maketitle
\tableofcontents

\section{Other models for magnetism in S\tlc{r}$_3$I\tlc{r}$_2$O$_7$}
Previous work on Sr$_3$Ir$_2$O$_7$ described the magnetic properties of the material using Heisenberg models, which project out any active charge degrees of freedom \cite{Kim2012_2, Moretti2015evidence, Gretarsson2016two, Hogan2016disordered, Li2020symmetry}. The magnetic dispersion was then calculated either under the linear spin-wave approximation or using bond-operator theory for a square lattice of dimers  \cite{Kim2012_2, Moretti2015evidence, Hogan2016disordered}. Although both models can be refined to reproduce features of the experimental spectra, it is notable that large numbers of unconstrained parameter are required and that both approaches lead to complications. 

The spin-wave treatment predicts a spectrum composed of two transverse optical and acoustic magnetic excitations with energies that should be equal upon reflection about the magnetic zone boundary, which is hard to reconcile with the measured spectra \cite{Kim2012_2, Moretti2015evidence}. In addition, multiple higher-order couplings that are barely weaker than the nearest neighbour interactions are required. These terms suggest that the $U/t_1\gg1$ limit, in which projecting the Hubbard model onto the Heisenberg model is valid, is not appropriate for Sr$_3$Ir$_2$O$_7$ \cite{Carter2013theory, Igarashi2014analysis}. Further, the model cannot account for the additional magnetic intensity at $q_c = 0.5$, as  physics beyond linear spin-wave theory is needed to model longitudinal quasiparticles. If spin-wave theory is extended to include a longitudinal response, a continuum and not an additional mode would be generated. More recently, a Raman scattering study proposed a spin-wave-theory based model for the magnetic excitations in Sr$_3$Ir$_2$O$_7$ \cite{Li2020symmetry}, but uses very different parameters to both of the other models and is inconsistent with both \gls*{RIXS} and inelastic neutron scattering spectra \cite{Fujita2021magnetic}.

The dimer approach does yield a longitudinal mode and a transverse mode, but it fails to predict the decay of the longitudinal mode away from $(0, 0)$ and $(0.5, 0.5)$. Another issue with this approach is that it requires magnetic interactions that are hard to physically justify as the ratio between the $c$-axis and $ab$-plane magnetic interactions should be much greater than one for this model to be valid, and to adequately reproduce the data \cite{Moretti2015evidence, Hogan2016disordered}. This is questionable because the in-plane Ir-Ir bonds in Sr$_3$Ir$_2$O$_7$ are actually shorter than the $c$-axis ones. The parameters required to make the dimer-bond-operator theory reproduce the observed magnetic excitation energy are also very different to those required in the spin-wave treatment even though both are derived from Heisenberg models. 

In contrast, the bilayer Hubbard model described in our work fits the data with more physically intuitive values. We employ a nearest-neighbour hopping ten times larger than the next-nearest-neighbour term, and a approximate one-to-one correspondence between intra- and interlayer hopping that is in line with the respective Ir-Ir bond lengths. The \gls*{RPA} also predicts the temperature dependence of the magnetic excitation satisfactorily (Fig.~\ref{fig:suppTdep}), providing evidence for the condensation of the excitonic longitudinal mode at $T_{\text{N}}$. This implies a close-by \gls*{QCP} which is indeed predicted by the theoretical model~\cite{Suwa21}, therefore clarifying the microscopic nature of the material. 

For completeness, we comment upon a further mechanism to generate magnetic longitudinal modes. This involves systems tuned to a \gls*{QCP} between states with spin 1 and spin 0 local moments in Mott insulators and can be realized by a subtle balance between non-cubic crystal field and moderate \gls*{SOC} \cite{Khaliullin2013excitonic}. This intriguing state has been realized in Ca$_2$RuO$_4$ \cite{Jain2017higgs}, but has minimal relevance to Sr$_3$Ir$_2$O$_7$ since it requires a d$^4$ electronic configuration. 

\section{Mean field and random phase approximations}
We modeled the magnetic excitation of Sr$_3$Ir$_2$O$_7$ using a half-filled bilayer Hubbard Hamiltonian $\mathcal{H}=-\mathcal{H}_{{\rm K}}+\mathcal{H}_{{\rm I}}$,
with $\mathcal{H}_{{\rm I}}\!=U\sum_{\bm{r}}n_{\bm{r}{\uparrow}}n_{\bm{r}{\downarrow}}$
and 
\begin{eqnarray}
\!\!\!\mathcal{H}_{{\rm K}}\!=\sum_{{\bm{r}},\bm{\delta}_{\nu}}t_{\nu}{\bm{c}}_{\bm{r}}^{\dagger}{\bm{c}}_{{\bm{r}}+{\bm{\delta}}_{\nu}}\!\!+t_{z}\!\!\sum_{{\bm{r}}_{\bot}}\!{\bm{c}}_{({\bm{r}}_{\bot},1)}^{\dagger}e^{i\frac{\alpha}{2}\epsilon_{\bm{r}}\sigma_{z}}{\bm{c}}_{({\bm{r}}_{\bot},2)}\!\!+\!{\rm H.c.},\label{eq:H_bl}
\end{eqnarray}
where $t_{\nu},t_{z},\alpha,U\in\mathbb{R}$ as defined in the main text. The system has an easy
$z$-axis spin anisotropy for $\alpha\neq0$, and the ground state
can have Néel ordering, $\langle S_{\bm{r}}^{\mu}\rangle=(-1)^{\gamma_{\bm{r}}}M\delta_{\mu z}$,
where $\gamma_{\bm{r}}=(1+\epsilon_{\bm{r}})/2$, ${S}_{\bm{r}}^{\mu}={1/2}c_{\bm{r}}^{\dagger}\sigma^{\mu}c_{\bm{r}}$
$(\mu=x,y,z)$ and $M$ is the magnetization. Following Ref.~\cite{Suwa21},
a mean-field decoupling of $\mathcal{H}_{{\rm I}}$ leads to ${\cal H}^{{\rm MF}}=\sum_{\bm{k}\in{\rm BZ}}{\bm{c}}_{\bm{k}}^{\dagger}{\cal H}_{\bm{k}}^{{\rm MF}}{\bm{c}}_{\bm{k}}^{\;}$
with 
\begin{align}
{\cal H}_{\bm{k}}^{{\rm MF}} & =\left(\begin{array}{cc}
\epsilon_{\bm{k}}^{(2)}+UM\sigma_{z} & \epsilon_{\bm{k}}^{(1)}-t_{z}\cos(k_{z})e^{-i\frac{\alpha}{2}\sigma_{z}}\\
\epsilon_{\bm{k}}^{(1)}-t_{z}\cos(k_{z})e^{i\frac{\alpha}{2}\sigma_{z}} & \epsilon_{\bm{k}}^{(2)}-UM\sigma_{z}
\end{array}\right),
\end{align}
where ${\bm{c}}_{\bm{k}}\equiv(c_{\mathcal{A}\uparrow,\bm{k}},c_{\mathcal{A}\downarrow,\bm{k}},c_{\mathcal{B}\uparrow,\bm{k}},c_{\mathcal{B}\downarrow,\bm{k}})^{T}$
is the Fourier transformed operator, and 
\begin{align}
\epsilon_{\bm{k}}^{(1)} & =-2t_{1}\left(\cos\frac{k_{1}+k_{2}}{2}+\cos\frac{k_{1}-k_{2}}{2}\right),\\
\epsilon_{\bm{k}}^{(2)} & =-2t_{2}\left(\cos k_{1}+\cos k_{2}\right),
\end{align}
and ${\bm{k}}\equiv({\bm{k}}_{\bot},k_{z})$ a wavevector in the first Brillouin
zone (BZ) \emph{i.e.} ${\bm{k}}_{\bot}=k_{1}{\bm{b}}_{1}^{\prime}+k_{2}{\bm{b}}_{2}^{\prime}$,
with ${\bm{b}}_{1}^{\prime}=(1/2,-1/2)$, ${\bm{b}}_{2}^{\prime}=(1/2,1/2)$,
{$k_{1},k_{2}\in[0,2\pi)$} and $k_{z}=0,\pi$. ${\cal H}_{\bm{k}}^{{\rm MF}}$ is diagonalized by a $4\times4$
unitary matrix $U(\bm{k})$, yielding 
\begin{eqnarray}
c_{\gamma\sigma,\bm{k}} & = & \sum_{n}U_{(\gamma\sigma),n}(\bm{k})\psi_{n,\bm{k}},
\end{eqnarray}
where $n\equiv(s,\sigma)$ with $s=\pm$, $\sigma=\uparrow,\downarrow$,
and each column of $U(\bm{k})$ is an eigenvector of ${\cal H}_{\bm{k}}^{{\rm MF}}$:
\begin{eqnarray}
X_{s\uparrow}(\bm{k})\!=\!\left(\begin{array}{c}
x_{\uparrow\bm{k}}\sqrt{\frac{1+sz_{\uparrow\bm{k}}}{2}}\\
0\\
s\sqrt{\frac{1-sz_{\uparrow\bm{k}}}{2}}\\
0
\end{array}\right),\;X_{s\downarrow}(\bm{k})\!=\!\left(\begin{array}{c}
0\\
x_{\downarrow\bm{k}}\sqrt{\frac{1-sz_{\downarrow\bm{k}}}{2}}\\
0\\
s\sqrt{\frac{1+sz_{\downarrow\bm{k}}}{2}}
\end{array}\right),\quad\ 
\end{eqnarray}
where 
\begin{eqnarray}
x_{\sigma\bm{k}} & = & \frac{b_{\sigma\bm{k}}}{\rvert b_{\sigma\bm{k}}\rvert},\;\;z_{\sigma\bm{k}}=\frac{\delta}{\sqrt{\delta^{2}+\rvert b_{\sigma\bm{k}}\rvert^{2}}},\;
b_{\sigma\bm{k}}= \epsilon_{\bm{k}}^{(1)}-t_{z}\cos(k_{z})e^{-i\sigma\frac{\alpha}{2}}
\end{eqnarray}
and $\delta=UM$. The corresponding eigenenergy, $\varepsilon_{s\sigma}(\bm{k})=\epsilon_{\bm{k}}^{(2)}+s\sqrt{\delta^{2}+\rvert b_{\sigma\bm{k}}\rvert^{2}},$
is independent of the spin flavour, $\rvert b_{\sigma\bm{k}}\rvert^{2}=b_{\bm{k}}^{2}$
and $z_{\sigma\bm{k}}\equiv z_{\bm{k}}$. The order parameter is determined by
solving the self-consistent equation:
\begin{align*}
\left(-1\right)^{\gamma}M & =\frac{1}{{\cal N}_{u}}\sum_{\bm{r}\in\gamma}\langle S_{\bm{r}}^{z}\rangle\\
 & =\frac{1}{2{\cal N}_{u}}\sum_{\bm{r}\in\gamma}\sum_{\sigma}\sigma\langle c_{\gamma\sigma,\bm{r}}^{\dagger}c_{\gamma\sigma,\bm{r}}\rangle\\
 & =\frac{1}{2{\cal N}_{u}}\sum_{\bm{k}}\sum_{\sigma}\sigma\langle c_{\gamma\sigma,\bm{k}}^{\dagger}c_{\gamma\sigma,\bm{k}}\rangle\\
 & =\frac{1}{2{\cal N}_{u}}\sum_{\bm{k}}\sum_{\sigma}\sigma\sum_{n}\left|U_{(\gamma\sigma),n}(\bm{k})\right|^{2}f_{n\bm{k}}\\
 & =\left(-1\right)^{\gamma}\frac{1}{2{\cal N}_{u}}\sum_{\bm{k}}z_{\bm{k}}\left(f_{-\bm{k}}-f_{+\bm{k}}\right),
\end{align*}
where ${\cal N}_{u}$ is the number of unit cells and $f_{s\bm{k}}=\frac{1}{1+e^{\beta\left(\varepsilon_{s}(\bm{k})-\mu\right)}}$
is the Fermi distribution function. We found the self-consistent solution
$\delta$ and $\mu$ under the condition the electron density satisfies
\begin{equation}
2n=\frac{1}{2N_{u}}\sum_{\bm{k},n}f_{n\bm{k}},\label{eq:cond_n}
\end{equation}
where $n$ is the electron density of the system. At half-filling ($n=1/2$)
and zero temperature, the critical value of $U$ can be obtained by
\begin{align}
\frac{1}{U_{{\rm c}}} & =\frac{1}{2{\cal N}_{u}}\sum_{\bm{k}}\left.\frac{\partial}{\partial\delta}z_{\bm{k}}\right|_{\delta=0} =\frac{1}{2{\cal N}_{u}}\sum_{\bm{k}}\frac{1}{\left|b_{\bm{k}}\right|}.
\end{align}
In the ordered phase ($U>U_{{\rm c}}$), the critical temperature
is obtained by solving the equation
\begin{align}
\frac{1}{U} & =\frac{1}{2{\cal N}_{u}}\sum_{\bm{k}}\left.\frac{\partial}{\partial\delta}\left(z_{\bm{k}}\left(f_{-\bm{k}}-f_{+\bm{k}}\right)\right)\right|_{\delta=0} =\frac{1}{2{\cal N}_{u}}\sum_{\bm{k}}\frac{f_{-\bm{k}}-f_{+\bm{k}}}{\left|b_{\bm{k}}\right|}
\end{align}
together with Eq. (\ref{eq:cond_n}) for $\delta=0$. The magnetic susceptibilities of the transverse
and the longitudinal modes within the \gls*{RPA} are
given by 
\begin{eqnarray}
\chi^{+-}(\bm{q},i\omega_{n}) & = & \frac{1}{\tau^{0}-U\chi_{0}^{+-}(\bm{q},i\omega_{n})}\chi_{0}^{+-}(\bm{q},i\omega_{n}),\label{eq:Trans_pm}\\
\chi^{zz}(\bm{q},i\omega_{n}) & = & \frac{1}{\tau^{0}-\frac{U}{2}\chi_{0}^{zz}(\bm{q},i\omega_{n})}\chi_{0}^{zz}(\bm{q},i\omega_{n}),\label{eq:rpa_L}
\end{eqnarray}
respectively, where $\tau_{0}$ is the $2\times2$ identity matrix.
Here $\chi^{+-}$ and $\chi^{zz}$ are $2\times2$ matrices in the
sublattice space, while $\chi_{0}^{+-}$ and $\chi_{0}^{zz}$ are
the bare magnetic susceptibilities~\cite{Suwa21}.
In the calculations of the spectra shown in Fig.~\ref{fig:suppTdep}, we set the broadening factor to $\eta=10^{-4}$~eV. The dynamical spin structure factors in Figs.~3c and d of the main manuscript are shown after convolution with the experimental resolution.

\section{Illustration of the exciton}

\begin{figure}[H]
\begin{center}
\includegraphics[width=.7\linewidth]{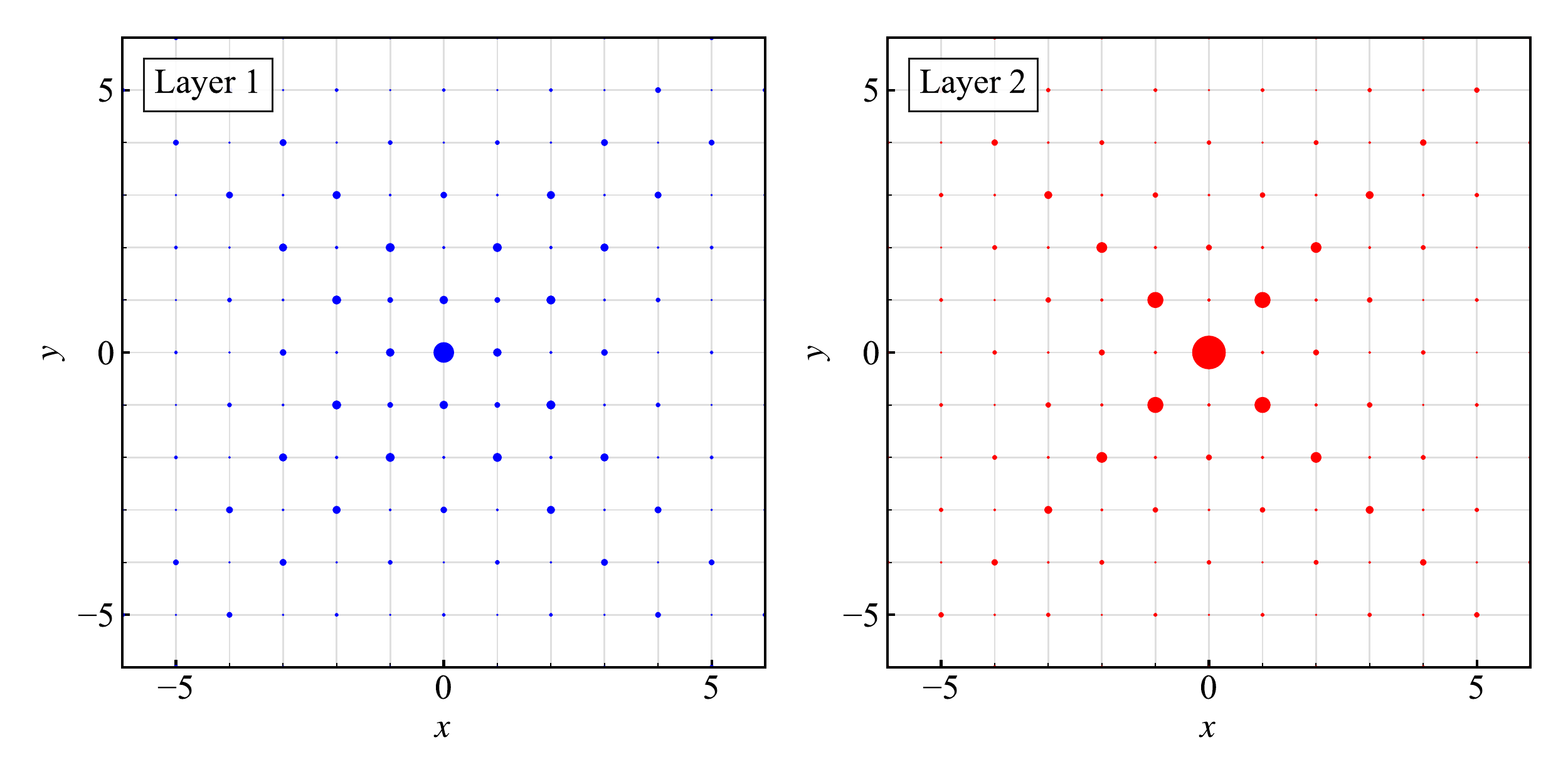} %
\caption{\textbf{Illustration of the exciton.} Left and right panels show the real-space Ir lattice for the two layers that make up the Sr$_3$Ir$_2$O$_7$ bilayer structure. The area of the circles at each lattice point is proportional to the modulus squared of the exciton wavefunction under the condition that a spin down hole is fixed at the origin of layer 1.  The left panel shows an enhanced probability of a spin-down electron surrounding the reference hole in layer 1, and the right panel shows an enhanced occupation living above the hole in layer 2. $x$ and $y$ represent the two orthogonal direction within the bilayer in units of the Ir-Ir spacing.}

\label{fig:exciton_wavefunction}
\end{center}
\end{figure}

In an antiferromagnetic excitonic insulator, the relevant bound electron-hole pairs are magnetic modes that are not directly visible to charge-sensitive spectral probes. We can, however, plot the spatial distribution of the electron and hole to illustrate their real space configuration, based on our theory, which is validated by its close agreement with our measurements of the magnetic dispersion and the observed charge gap. As seen in Fig.~\ref{fig:exciton_wavefunction} the exciton is made up of a specific spin and charge configuration distributed between the two layers and dispersed over a few lattice constant laterally. In Fig.~\ref{fig:exciton_wavefunction} the state is represented in terms of electron/hole population with respect to the average band filling of 1 electron per orbital, as pairing between spin-down holes with spin-down electrons, which is equivalent to pairing electrons with opposite spin.

\section{Excitonic longitudinal mode condensation at $T_\text{N}$}
\begin{figure}[H]
\includegraphics[width=1\linewidth]{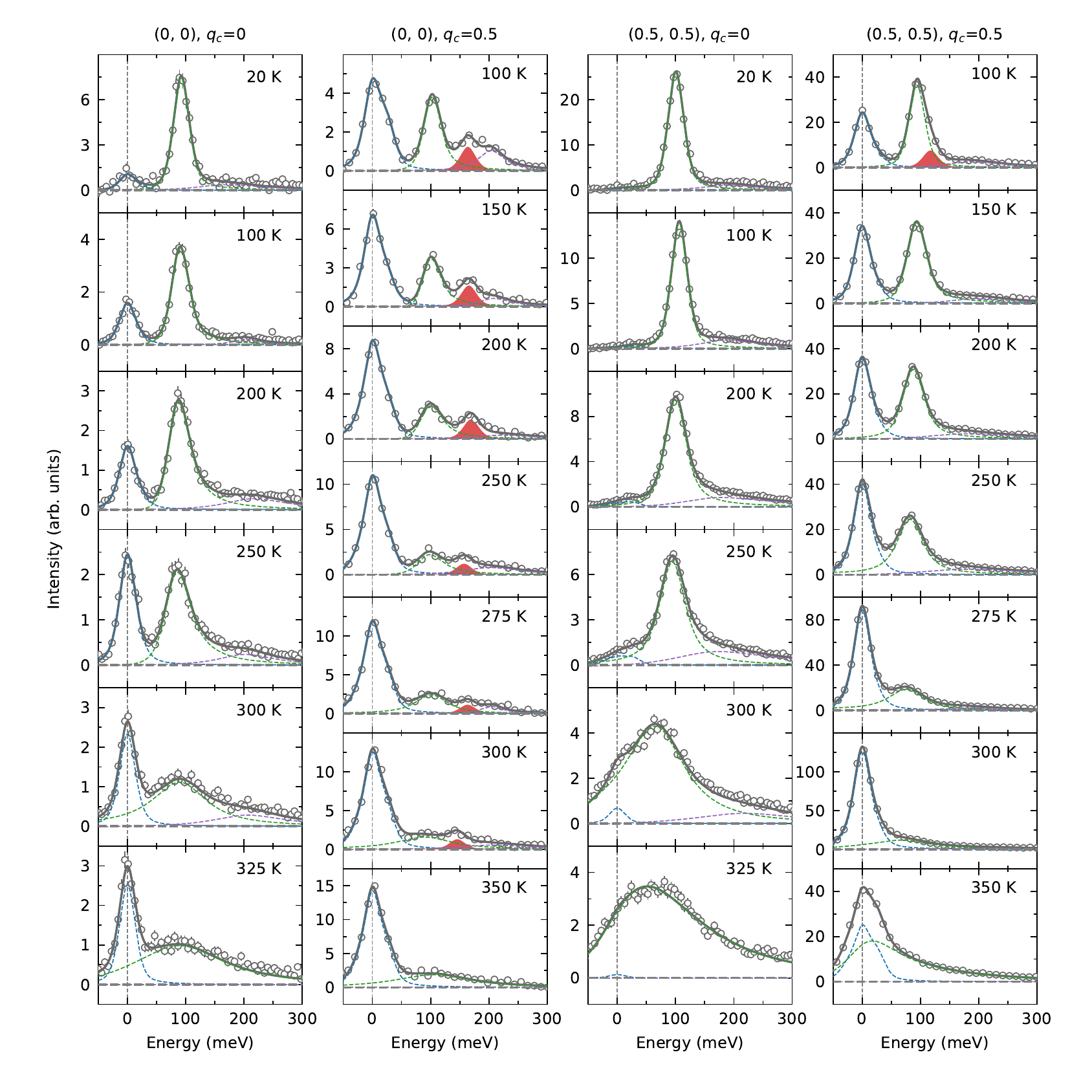} %
\caption{\textbf{Temperature dependence of energy-loss spectra at (0, 0) and (0.5, 0.5).}  \Gls*{RIXS} spectra at (0, 0) and (0.5, 0.5) with $q_c=0$ and 0.5 for specific temperatures. The black circles represent the data and dotted lines outline the different components of the spectrum, which are summed to produce the solid line representing the total spectrum. The fitting procedure is described in detail in the Methods section of the main manuscript. Error bars are determined using Possonian statistics. The results support the longitudinal mode condensation at $T_\text{N}$.}
\label{fig:suppltdep}
\end{figure}

The involvement of the excitonic longitudinal mode in the formation of ground-state magnetism was studied using thermal fluctuations. Figures 4a-d of the main manuscript show the temperature dependence of the magnetic excitation spectra at (0, 0) and $(0.5, 0.5)$ for $q_c=0$ and 0.5, which were generated from temperature dependent energy-loss spectra such as shown in Fig.~\ref{fig:suppltdep}.

\begin{figure}[H]
\begin{center}
\includegraphics[width=0.5\linewidth]{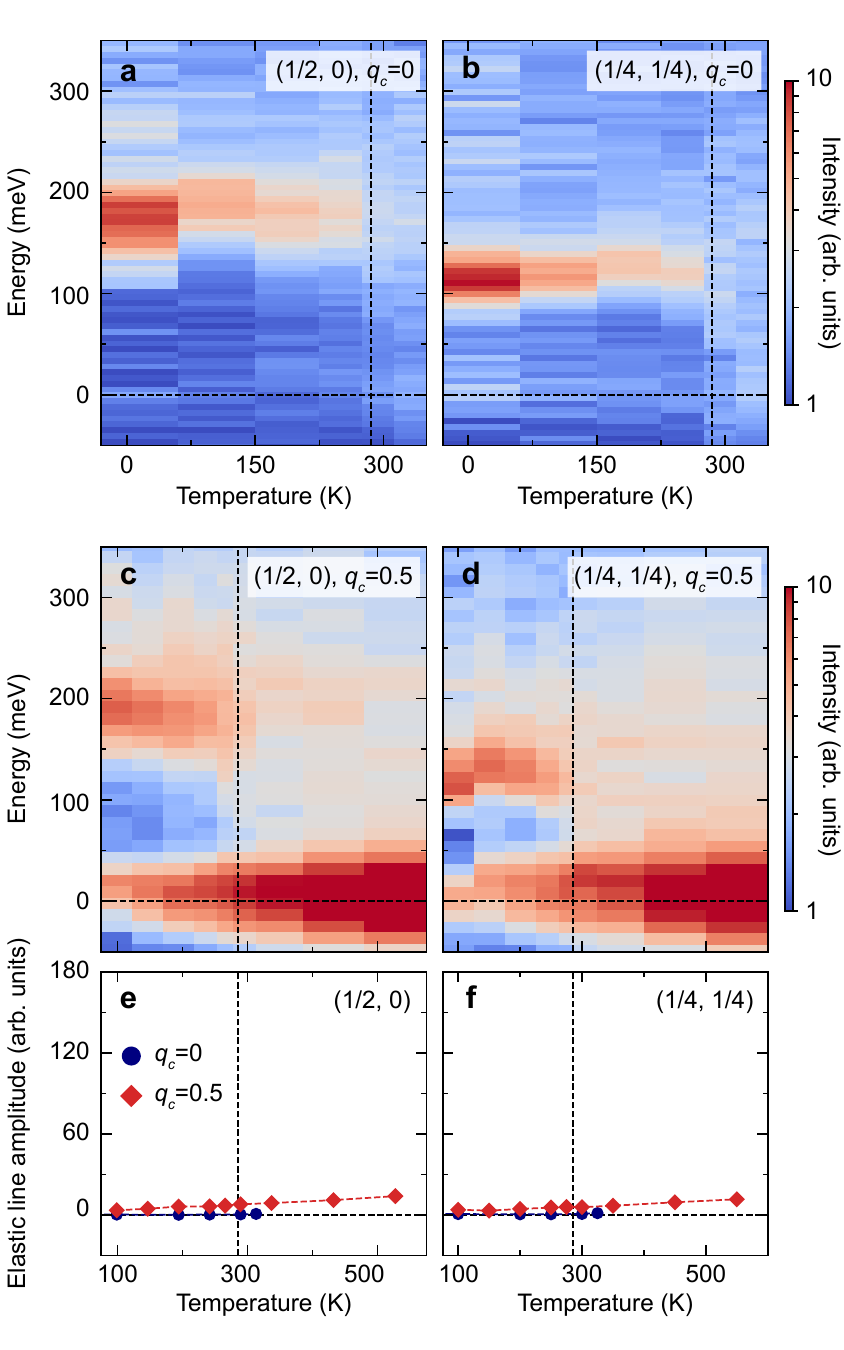} %
\caption{\textbf{Additional temperature-dependent energy-loss spectra.} \textbf{a-d} Temperature dependence of the Sr$_3$Ir$_2$O$_7$ excitation spectrum at $(0.5, 0)$ and $(0.25, 0.25)$  for $q_c = 0$ and 0.5. The color bar indicates the strength of the magnetic excitations. \textbf{e, f} Quasi-elastic intensity as function of temperature for $q_c = 0$ and 0.5 in blue and red, respectively. The data reveal the expected monotonic enhancement of the quasi-elastic intensity with increasing temperature.}
\label{fig:suppTdepexperiment}
\end{center}
\end{figure}

The data reveal a substantial softening of the excitonic longitudinal mode, while only minimal detectable softening is observed in the transverse channel. This is supported by further temperature dependent energy-loss spectra at $(0.5, 0)$ and $(0.25, 0.25)$ shown in Figs.~\ref{fig:suppTdepexperiment}a-d. In agreement with the temperature-dependence of the quasi-elastic intensity at (0, 0) (Fig.~4e in the main manuscript), a monotonic enhancement is also observed at $(0.5, 0)$ and $(0.25, 0.25)$ (Figs.~\ref{fig:suppTdepexperiment}e-f).

\begin{figure}[H]
\begin{center}
\includegraphics[width=0.5\linewidth]{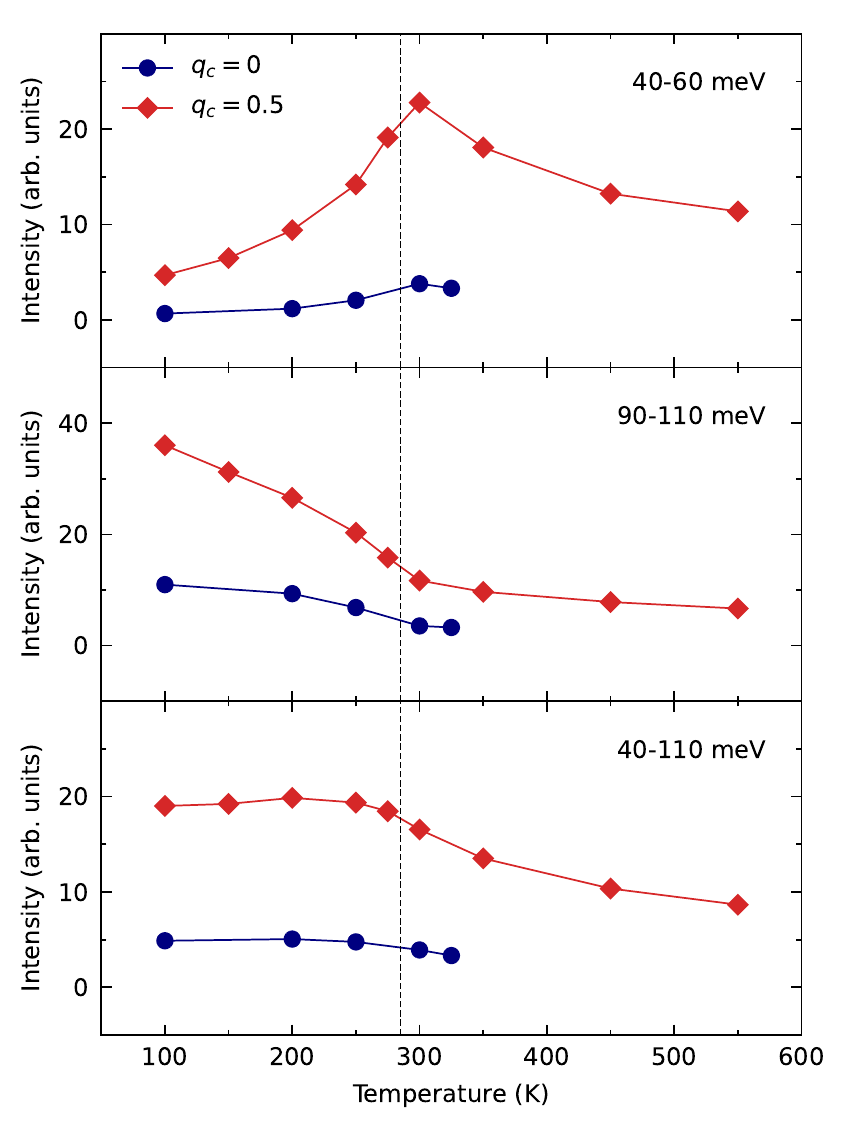} 
\caption{\textbf{Temperature dependence of the low-energy excitation spectrum.} \textbf{a-d} Temperature dependence of the low-energy excitation spectrum integrated over an energy range of 40-60, 90-110 and 40-110 meV at (0.5, 0.5) $q_c = 0$ and 0.5. Although the damping effect of the excitonic longitudinal mode makes it difficult to directly attribute the softening to the longitudinal mode, a mode condensation at (0.5, 0.5) $q_c = 0.5$ is further supported by the enhancement of the low-energy quasi-elastic intensity.}
\label{fig:supplowE}
\end{center}
\end{figure}

As explained in the main text, increasing temperature broadens the spectra considerably, so we leverage the symmetry properties of the modes to unveil the detailed mode behavior. The transverse mode energy is predicted to be independent of $q_c$ and we can study this mode in isolation at $q_c=0$ and see that it shows minimal softening. At $q_c=0.5$, however, the transverse and longitudinal modes are present and considerable softening is observed, but since the transverse mode has minimal softening, we can assign this softening to arise predominately from the longitudinal mode. Although the damping effect of the excitonic longitudinal mode makes it difficult to directly attribute the softening to the longitudinal mode, a mode condensation at (0.5, 0.5) $q_c = 0.5$ is further supported by the enhancement of the low-energy excitation spectrum (0.5, 0.5) $q_c = 0.5$ that is absent at $q_c = 0$ (see Fig.~\ref{fig:supplowE}).  

\begin{figure}[H]
\includegraphics[width=1\linewidth]{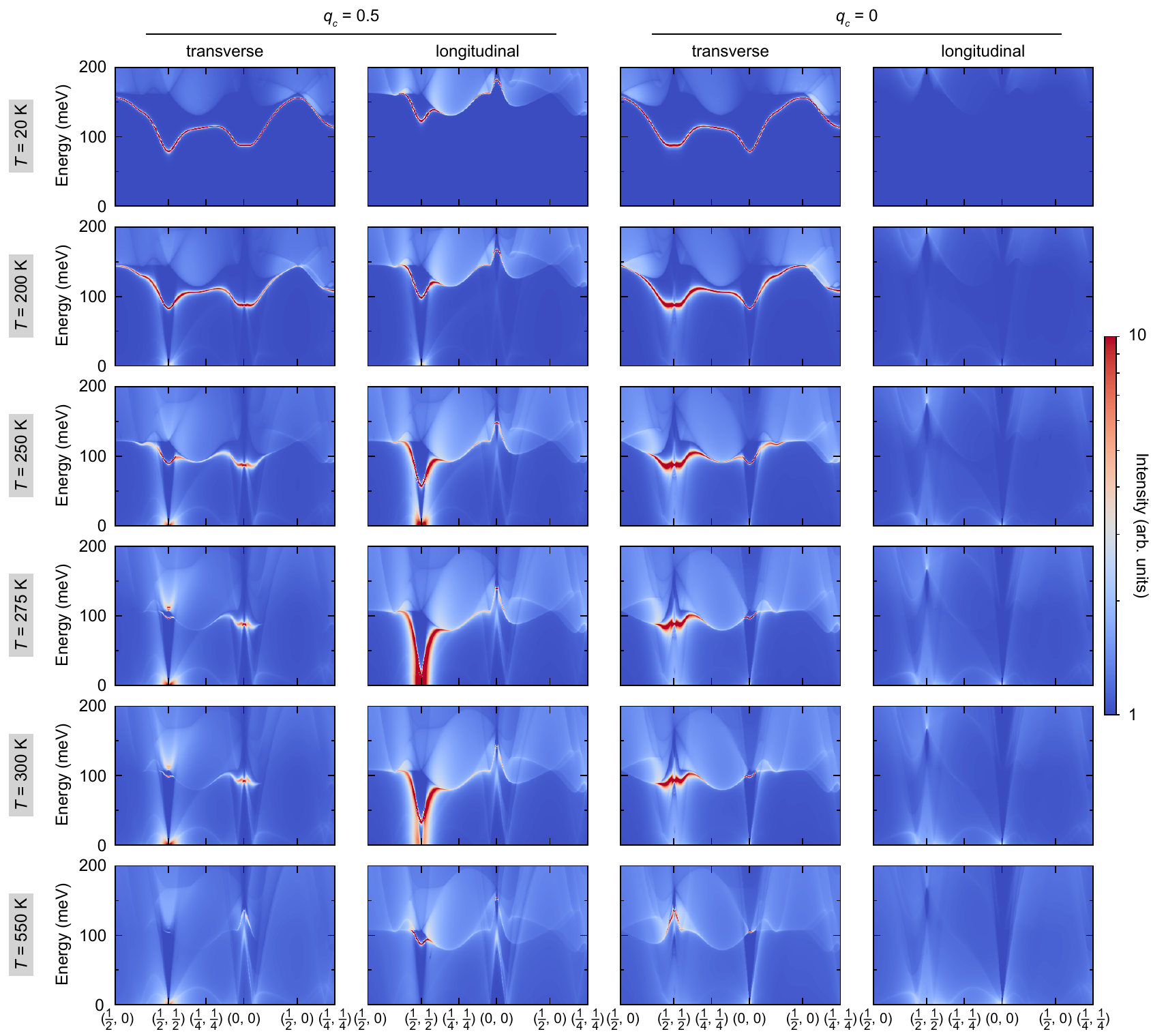} %
\caption{\textbf{Theoretical prediction of excitonic mode condensation at $T_\text{N}$.} Theoretical calculation of the transverse and longitudinal magnetic excitation spectrum at $q_c = 0$ and 0.5 as function of temperature. The color bar indicates the strength of the magnetic excitations.}
\label{fig:suppTdep}
\end{figure}

The temperature dependence of the transverse and longitudinal magnetic excitation spectrum was calculated using a \gls*{RPA} in the thermodynamic limit (the Hubbard Hamiltonian is described in the Methods section of the main manuscript). Figure~\ref{fig:suppTdep} displays the predicted dynamic structure factor at $q_c = 0$ and 0.5 for $T$ = 20, 200, 250, 275, 300 and 550 K. The color bar indicates the strength of the magnetic excitations. The results show that upon increasing temperature the transverse mode has only minimal detectable softening, which is expected in view of the Ising nature of magnetism. This is in strong contrast with the temperature dependence of the longitudinal mode at $q_c=0.5$. At $T=20$~K the exctonic longitudinal mode is well-defined around (0.5, 0.5) and (0, 0), and dissolves into the electron-hole continuum at wavevetors away from these reciprocal-lattice positions. Increasing temperature yields a continuous softening and ultimately to the condensation of the mode at $T_\text{N}=285$~K. This shows that the excitonic longitudinal mode establishes the magnetic long-range order in Sr$_3$Ir$_2$O$_7$.

Above $T_\text{N}$ our calculations reveal remnant longitudinal intensity, which can be assigned to preformed exciton pairs. This classifies the phase transition as a Bose-Einstein condensation, opposed to Bardeen-Cooper-Schrieffer-like transition. The Bose-Einstein nature can be traced back to the \gls*{SOC} of the theoretical model shown in the Methods section of the main manuscript. The finite \gls*{SOC} included in the interlayer hopping establishes a narrow-band insulator even at $U=0$. Thus, preformed excitons exist in the paramagnetic state at  $U<U_c$, and $U>U_c$ for $T>T_\text{N}$.

\section{Minimal contribution of critical scattering to quasi-elastic intensity}
\begin{figure}[H]
\begin{center}
\includegraphics[width=.7\linewidth]{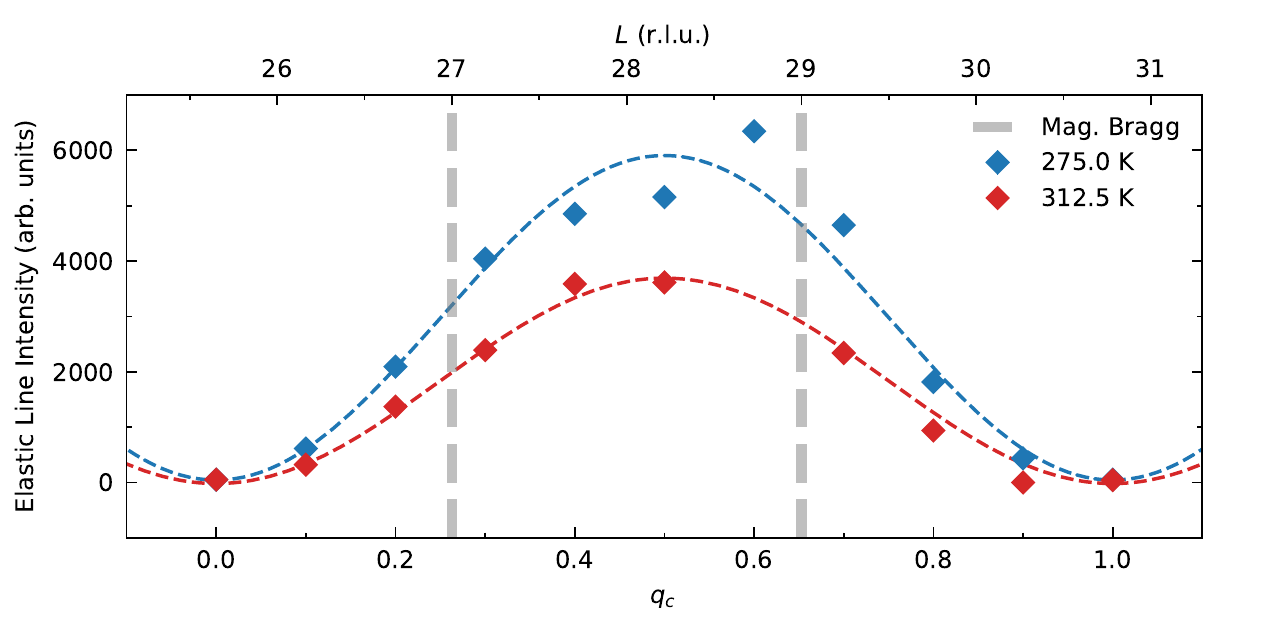} %
\caption{\textbf{Origin of quasi-elastic intensity near $T_\text{N}$.} The $c$-axis momentum dependence of the quasi-elastic intensity at $(0.5, 0.5)$ is plotted at two temperatures near $T_\text{N}$. The overall tend in intensity follows a $\sin^2(\pi q_c)$ dependence illustrated by the dotted line, as expected for intensity that arises from longitudinal mode softening. This is quite distinct from critical scattering from N\'{e}el order, which would be centered around the magnetic Bragg peaks shown as vertical gray lines.}
\label{fig:critical}
\end{center}
\end{figure}

In the main manuscript, we interpret the temperature-dependent \gls*{RIXS} spectra in terms of longitudinal mode softening. Another possible source of quasi-elastic scattering is critical scattering associated with the sample's N\'{e}el order. As mentioned in the main text, the bilayer separation in Sr$_3$Ir$_2$O$_7$ is incommensurate with the $c$-axis constant, so critical scattering from N\'{e}el order and quasi-elastic scattering from longitudinal mode softening can be distinguishing based on their $c$-axis momentum dependence. We plot the $q_c$-dependence of the $(0.5, 0.5)$ quasi-elastic line intensity at two temperatures in the vicinity of $T_\text{N}$ a function of $q_c$ in Fig.~\ref{fig:critical}. The results show that longitudinal mode softening dominates the measured intensity.

\section{Orbital currents}

As was predicted in Ref.~\cite{Halperin1968possible}, antiferromagnetic order in an excitonic insulator with finite \gls*{SOC} could in principle induce orbital antiferromagnetism associated with charge currents on plaquettes of the lattice. However, we have verified that this does not occur in Sr$_3$Ir$_2$O$_7$  by calculating the expectation value of the current operator around a plaquette. The only source of orbital antiferromagnetism in Sr$_3$Ir$_2$O$_7$  is the strong intra-atomic \gls*{SOC} that leads to a mixed spin-orbital character of the atomic magnetic moments.

\bibliographystyle{naturemag}
\bibliography{refs}